\documentstyle[floats,aps,epsf]{revtex}

\begin{document}
\title
{Superdeformation and clustering in $^{40}$Ca studied with Antisymmetrized Molecular 
Dynamics}

\author{Y. Kanada-En'yo and M. Kimura}

\address{Yukawa Institute for Theoretical Physics, Kyoto University,\\
Kyoto 606-8502, Japan}

\maketitle
\begin{abstract}
Deformed states in $^{40}$Ca are 
investigated with a method of antisymmetrized molecular dynamics.
Above the spherical ground state, rotational bands arise from 
a normal deformation and a superdeformation 
as well as an oblate deformation.
The calculated energy spectra and $E2$ transition strengths
in the superdeformed band reasonably agree to the experimental data of the 
superdeformed band starting from the $0^+_3$ state at 5.213 MeV.
By the analysis of single-particle orbits, 
it is found that the superdeformed state has particle-hole nature
of an $8p$-$8h$ configuration.
One of new findings is parity asymmetric structure 
with $^{12}$C+$^{28}$Si-like clustering in the superdeformed band.
We predict that $^{12}$C+$^{28}$Si molecular bands may be built 
above the superdeformed band due to the excitation of inter-cluster
motion. They are considered to be higher nodal states of the 
superdeformed state.
We also suggest negative-parity bands caused by the parity asymmetric
deformation.
\end{abstract}

\noindent

\section{Introduction}
In the spectra of $^{40}$Ca, the existence of low-lying deformed states has
long been known since 1960s\cite{Macdonald}. The coexistence of the
spherical and various deformed states in the doubly magic nuclei
$^{40}$Ca has 
attracted a great interest. In the positive parity states below 7 MeV,
besides the spherical ground state, the existence of two deformed
rotational bands built on the $0^+_2$ state at 3.35 MeV and the $0^+_3$
state at 5.21 MeV is known experimentally. The structure of these
bands has been often discussed in relation to 
many-particle many-hole states of nuclei in this mass region for a long
time. Grace and Green \cite{Gerace67,Gerace69} firstly suggested that
the former band is dominantly a $4p$-$4h$ state, while
the latter band with a large deformation is understood as an $8p$-$8h$
state.  

The member states of the first $K^\pi$=$0^+$ rotational band built on
the $0^+_2$ state are known to be strongly populated by
$\alpha$-transfer reactions and to have the large $\alpha$ spectroscopic 
factors \cite{Betts77,Fortune79,Yamaya94,Yamaya98}. 
The $\alpha$+$^{36}{\rm Ar}$
cluster structure of this band has been studied theoretically 
based on semi-microscopic cluster model 
calculations\cite{Ohkubo88,Sakuda94,Sakuda98}.  

The states in the second $K^\pi$=$0^+$ band have been observed
in the experimental work searching for the $8p$-$8h$ states with 
$^{32}$S($^{12}$C,$\alpha$)$^{40}$Ca reactions, which were performed by
Middleton {\em et al.} in 1972 \cite{Middleton72}. 
Due to the strong population in the multi-nucleon transfer data and
the strong $E2$ transitions \cite{Macdonald},
the $0^+$(5.21 MeV), $2^+$(5.63 MeV) and $4^+$
(6.54 MeV) have been thought to belong to the superdeformed band 
with a dominant $8p$-$8h$ configuration.
Recently, by using the GAMMASPHEAR
array detectors, the level structure of the
deformed bands in $^{40}$Ca has been explored and many excited states 
up to very high spins have been discovered \cite{Ideguchi01,Chiara03}. 
Consequently, the existence of the superdeformed band
of $^{40}$Ca built on the $0_3^+$(5.21 MeV) has been experimentally well 
established.

In contrast to the experimental situation, theoretical interpretation of
the superdeformed band is not well established and is still 
complicated. With the symmetry restricted Skyrme-Hartree-Fock(HF)
calculations, Zheng {\em et al.} studied the energy systematics of
$n$particle-$n$hole($np$-$nh$) states in $^{40}$Ca\cite{Zheng88,Zheng90}. 
They suggested an $8p$-$8h$ deformed states as a local minimum of 
the energy as a function of $n$. They also predicted highly 
deformed $12p$-$12h$ states
in a high excitation energy region. However, the recent mean-field
calculations suggest that the $8p$-$8h$ configuration becomes
unstable, especially when the model space is extended.
Based on the three-dimensional(3d) HF calculations without any
assumption of the spatial symmetry, Inakura {\em et al.} studied the
energy surface of $^{40}{\rm Ca}$ as the function of the quadrupole
deformation. They have shown that the local minimum of the superdeformed
$8p$-$8h$ configuration on the energy curve becomes quite shallow or even 
disappear due to the $\gamma$ and octupole deformations 
depending on the effective interaction, 
though the $12p$-$12h$ state and an oblate state 
remain to be local minima in any of used effective interactions. 
They have pointed out that the
extreme softness  of the superdeformed $8p$-$8h$ configuration against the
octupole deformation ($Y_{30}$ and $Y_{31}$). 
The instability of the
$8p$-$8h$ configuration and the stability of the $12p$-$12h$ 
and oblate solutions in $^{40}$Ca had been also reported by 
Leander and Larsson\cite{Leander75} based on 
the macroscopic-microscopic model calculations.  
Further calculations beyond mean-field theory were performed by
Bender {\em et al.} with GCM calculations in HF+BCS approach 
\cite{Bender03}. In their results, 
the superdeformed band is not
dominated by a certain $np$-$nh$ configuration but it contains various
configurations such as $4p$-$4h$, $6p$-$6h$, and $8p$-$8h$ states due to
the strong mixture between these configurations. 
Moreover, shell model calculations were performed
for description of the superdeformed bands\cite{Poves04}.
Although the level structure and quadrupole moments of the superdeformed band 
are described by the spherical shell model with 
the fixed $8p$-$8h$ configurations, however, when the model space 
is extended to include other particle-hole configurations, 
the quadrupole moments are much underestimated due to the configuration mixing.

These facts may imply that a stable solution for the intrinsic state of
the superdeformation is hardly obtained in the modern calculations with
mean-field approaches. It seems to be somehow inconsistent with the
experimental energy spectra of the superdeformation in which no large
deviation from the rigid rotor model is seen
\cite{Ideguchi01,Chiara03}. Therefore, it is natural to expect that
there might be some mechanism beyond mean-field approaches to stabilize
the superdeformation. The cluster aspect is
one of the essential features for nuclear deformation even in 
$sd$-$pf$ shell nuclei together with the mean-field aspect.
Indeed, it has been already shown that two
kinds of nature, i.e. cluster and mean-field coexist 
in the deformed states in $sd$-shell 
nuclei \cite{AMDrev,Kimura-Ne20,Kimura-S32}.
We expect that the clustering can play an important role also 
in the formation of the superdeformed state of $^{40}{\rm Ca}$.
In case of $^{40}$Ca, we consider that the
softness against octupole deformations in the superdeformed state
suggested by Inakura {\em et al.}
can be associated with asymmetric di-cluster structure, though, to our 
knowledge, there has been no studies to investigate its cluster aspect.

We also note that some $sd$-shell nuclei have the remarkable 
cluster aspect, molecular resonances, whose properties 
have been investigated by cluster models. 
Recently, there are theoretical efforts to link 
the molecular resonances with low-lying deformed states
with full microscopic calculations.
It is suggested that the molecular resonances
and the low-lying deformed states in $sd$-shell 
are regarded to be a series of molecular bands. For example, 
in $^{32}$S system, it has been suggested that
$^{16}$O-$^{16}$O molecular resonances are 
built on the superdeformed band of $^{32}$S \cite{Kimura-S32} owing to
the excitation of the inter-cluster motion. Therefore, we consider that 
molecular resonances can be good probes to understand the cluster
aspect of the low-lying deformed states. In $^{40}$Ca system, there are
experimental implications of $^{28}$Si+$^{12}$C molecular resonances in
the low-energy  fusion cross section\cite{Racca83} and also elastic 
scattering data\cite{Ost79}, where distinct structures have been found
in the energy region  $E_{\rm cm}$=20-30 MeV above the
$^{28}$Si+$^{12}$C threshold. It is interesting to look into
possible appearance of  molecular resonances and its relation to the
superdeformation in $^{40}$Ca. 

In this paper, we study deformed states in $^{40}$Ca with antisymmetrized
molecular dynamics(AMD). The method of AMD has been proved to be useful
in description of shape coexistence as well as cluster structure 
in $sd$-shell nuclei
\cite{AMDrev,Kimura-Ne20,Kimura-S32,Kanada-Ne20,Kimura01,Kimura-Ne30,Enyo-n14,Taniguchi-Ne20}.
One of advantages of this method is that both cluster aspect and
mean-field aspect can be described within the AMD framework. In order to
study the coexistence of deformations in $^{40}$Ca and to investigate
the corresponding rotational bands,  we apply a constraint AMD
method. Parity projection, which is essentially important to describe
the parity asymmetric shape of nuclei, is performed before energy
variation,  while total-angular-momentum projection is  done after the
energy variation. We also  perform superposition of the wave functions
with different configurations obtained by the constraint AMD to obtain
better wave functions and to describe possible configuration mixing.  
Level scheme and $E2$ transition strengths are calculated from these
superposed wave functions. In the present work, we adopt the effective 
nuclear interaction  which contains finite-range 2-body and 3-body
forces proposed in Ref.\cite{ENYO-3fb}. This interaction is suitable for
describing the nuclear structure  properties such as binding energies
and radii over a wide mass number region from $^4$He to
$^{40}$Ca. Moreover, it is also useful to describe nucleus-nucleus
potential.  
These features can not be described by zero-range forces such as 
the Skyrme forces, and are superior points of  
the effective forces with finite-range three-body terms. 
We believe that they are essential to represent cluster aspect and 
to investigate properties of many-particle-many-hole 
states with large deformations.

This paper is organized as follows.
In section \ref{sec:formulation}, the formulation of AMD is 
briefly explained. 
In Sec.\ref{sec:results}, we show the calculated results of the energies,
quadrupole transitions and moments, and also the intrinsic structure.
We give discussions of molecular aspect and single-particle orbits
in Sec.\ref{sec:discuss}. 
Finally, a summary is given in Sec.\ref{sec:summary}.

\section{Formulation}\label{sec:formulation}
We here briefly explain the formulation of the present calculations.
For the details of the AMD formulation for nuclear structure studies, 
the readers are referred to Refs. 
\cite{AMDrev,ENYObc,ENYOsup}.

\subsection{Intrinsic wave function}
We start from the intrinsic wave function for a $A$-nucleon system 
which is expressed by a Slater 
determinant of single-particle wave packets,
\begin{eqnarray}\label{eq:AMD}
 |\Phi_{\rm AMD}\rangle &=& \frac{1}{\sqrt{A!}} {\cal{A}} \{
  \varphi_1,\varphi_2,...,\varphi_A \},\\
 \varphi_i({\bf r}_j) &=& \phi_i({\bf r}_j)\chi_i\eta_i,
\end{eqnarray}
where ${\cal A}$ is the antisymmetrizer, and $\varphi_i$ is the $i$th single-particle wave packet consisting
of the spatial part $\phi_i$, the spin part $\chi_i$ and the isospin
part $\eta_i$. The spatial part of the $i$th single-particle wave packet is
written by a Gaussian wave packet located at ${\bf X}_i$.
\begin{eqnarray}\label{eq:varphi}
  \phi_i({\bf r}_j) &=& \exp\bigl\{-\nu({\bf r}_j-{\bf X}_i)^2\bigr\},\\
 \chi_i &=& (\frac{1}{2}+\xi_i)\chi_{\uparrow}
 + (\frac{1}{2}-\xi_i)\chi_{\downarrow},\\
 \eta_i &=& {\rm proton\quad or \quad neutron}. 
\end{eqnarray}
Here, the center ${\bf X}_i$ of Gaussian is a complex number
and treated as a independent free parameter for each nucleon. 
The width parameter $\nu$ takes a common value
for all nucleons, and is chosen to be an optimum value for each nucleus. 
The orientation of the intrinsic spin is expressed by
a variational complex parameter $\xi_{i}$, and the isospin
function is fixed to be up(proton) or down(neutron). 
Thus, an AMD wave function
is specified by a set of variation parameters, ${\bf Z}\equiv 
\{{\bf X}_1,{\bf X}_2,\cdots, {\bf X}_A,\xi_1,\xi_2,\cdots,\xi_A \}$,
which are optimized by the energy variation.
As the variational wave function, we adopt the parity projected AMD 
wave function,
\begin{eqnarray}
 |\Phi^{\pm}_{\rm AMD}\rangle &=& \frac{1\pm P_x}{2} |\Phi_{\rm AMD}\rangle.
\end{eqnarray}
Thus the parity projection is performed 
before the variation. This is
essentially important to describe the parity asymmetric intrinsic
structure such as the asymmetric cluster structure as discussed in 
Refs.\cite{AMDrev,Marcos83}.

\subsection{Hamiltonian and constraint}

The Hamiltonian consists of the kinetic energy, nuclear and Coulomb
forces. In the present work, we use the effective nuclear force 
consisting of two-body and three-body forces. Then the Hamiltonian
is written as,
\begin{equation}
\hat H=\sum_i \hat t_i +\sum_{i<j} \hat v^{(2)}_{ij}+\sum_{i<j<k} 
\hat v^{(3)}_{ijk}-\hat T_{\rm cm},
\end{equation}
where $\hat t_i$ is a kinetic operator, 
$\hat v^{(2)}$ and $\hat v^{(3)}$ are two-body and three-body forces, 
respectively. $\hat T_{\rm cm}$ is the kinetic term for the center-of-mass
motion. In the AMD wave function, the energy of center-of-mass
motion is subtracted exactly.

We apply a constraint AMD method, where
we find the minimum energy solution in the AMD model space
under certain constraints. Since the particle-hole structure 
of the low-lying state is one of the main interests in the present work,
we adopt the constraint of
a total number of harmonic oscillator(H.O.) quanta,
\begin{equation}
\hat N_{\rm os}\equiv \sum _i \left[\frac{{\bf p}^2_i}{4\hbar^2\nu}
+\nu{\bf r}^2_i -\frac{3}{2}\right ],
\end{equation}
where the width parameter $\nu$ is taken to be the same value as that of the 
single-particle wave packets in Eq.(\ref{eq:varphi}).
This is the principal quantum number of the 
spherical harmonic oscillator with the oscillation number 
$\omega=2\hbar\nu/ m$($m$ is the nucleon mass).
In order to obtain the minimum energy state with the constraint
$\langle \hat N_{\rm os}\rangle = N_{\rm os}$, where $N_{\rm os}$ is a given
number, we perform energy variation for the expectation value ${\cal E}$ 
of the Hamiltonian,
\begin{equation}
{\cal E}\equiv 
\frac{\langle \Phi^\pm_{\rm AMD}|\hat H|\Phi^\pm_{\rm AMD}\rangle}
{\langle \Phi^\pm_{\rm AMD}|\Phi^\pm_{\rm AMD}\rangle},
\end{equation}
with respect to the variational parameters ${\bf Z}$ 
by the method of frictional cooling \cite{AMDrev}
under the constraint.
We note the AMD wave function obtained with the energy variation for
positive or negative parity state under the
constraint $\langle \hat N_{\rm os}\rangle = N_{\rm os}$ as
$\Phi_{\rm AMD}(N_{\rm os}^{+})$ or $\Phi_{\rm AMD}(N_{\rm os}^{-})$,
respectively. 

\subsection{superposition}

In the obtained wave function $\Phi_{\rm AMD}(N_{\rm os}^{(\pm)})$,
there exist local minimum solutions in the energy curve as a function
of $N_{\rm os}$. As shown later, those are usually the local minima 
also with respect to the deformation parameter $\beta$ and considered to
be approximate intrinsic wave functions for the corresponding 
deformed bands. In the calculation of level scheme and physical observables,
we choose several intrinsic wave functions 
$\Phi_{\rm AMD}(k)$ and superpose the spin-parity projected wave functions,
$P^J_{MK}\Phi^\pm_{\rm AMD}(k)$,
in order to satisfy the orthogonality among the calculated states
and to obtain better wave functions.
Here, $k$ indicates $N_{\rm os}^{(\pm)}$ and stands for the label for the 
obtained intrinsic wave function($\Phi_{\rm AMD}(k=N_{\rm os}^{(\pm)})$).
The wave function of the $n$th $J^\pm$ state is written as,
\begin{equation}
\Psi_n^{J\pm}=\sum_{k,K} c^{J^\pm_n}_K(k)P^J_{MK}\Phi^\pm_{\rm AMD}(k),
\end{equation}
where coefficients $c^{J^\pm_n}_K$ are determined by digonalizing the 
Hamiltonian and norm matrices:
\begin{eqnarray}
&\langle P^J_{MK'}\Phi^\pm_{\rm AMD}(k')|\hat H|P^J_{MK''}\Phi^\pm_{\rm AMD}(k'')\rangle \qquad {\rm and}\\
&\langle P^J_{MK'}\Phi^\pm_{\rm AMD}(k')|
P^J_{MK''}\Phi^\pm_{\rm AMD}(k'')\rangle.
\end{eqnarray}
Physical quantity for an operator $\hat O$ for the state
is calculated as, 
\begin{equation}
\langle \hat O \rangle=\langle\Psi_n^{J\pm}|\hat O|\Psi_n^{J\pm}\rangle.
\end{equation}
If we use an enough number of base wave functions, this procedure of the 
superposition is equivalent to the generator coordinate method
with respect to the generator coordinate $N_{\rm os}$. 
However, the number of the basis in the superposition is
10 at most because of huge computational time in the numerical calculations 
of the total 
angular momentum projection in the present work. It means that $\sum_{k}$ 
stands for the summation of discrete basis, 
and hence, the present calculations do not correspond to 
the complete GCM calculations.

\section{Results}\label{sec:results}

\subsection{Effective nuclear force}
We use an effective nuclear force which consists of central force,
spin-orbit and Coulomb forces. 
As the central force, we use F3B force\cite{ENYO-3fb}, 
which contains a finite-range three-body term.
The parameter set (2) of the F3B force 
is adopted. As for the spin-orbit force, we use the same interaction as
that adopted in Ref.\cite{ENYO-3fb}, the two-range Gaussian spin-orbit term
in G3RS force\cite{LS}. In the present calculations,
the strength parameter of the spin-orbit force is chosen to be 
(i)$u_{ls}\equiv u_1=-u_2=2500$ MeV
or (ii)$u_{ls}=1800$ MeV. 
Coulomb force is approximated by the sum of seven Gaussians.

\begin{figure}
\noindent
\epsfxsize=0.35\textwidth
\centerline{\epsffile{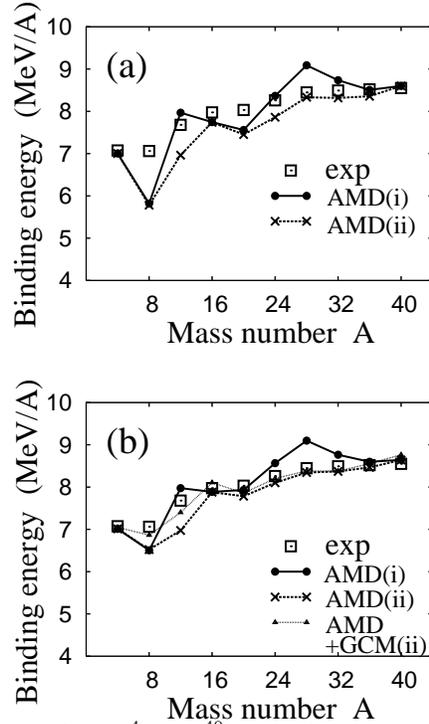}}
\caption{\label{fig:be}
The binding energies of 
$Z=N$ nuclei from $^4$He to $^{40}$Ca obtained by 
the simple AMD calculations without constraints.
The binding energies calculated by variation after parity 
projection but no total-angular momentum projection 
are shown in the upper panel(a), and the energies obtained by 
total-angular-momentum projection after the variation with parity
projection are plotted in
the lower panel(b).
Squares are the experimental binding energies.
Filled circles and crosses indicate the results by using the interaction(i)
 and (ii), respectively.
The results for the interaction(ii) in Ref.\protect\cite{ENYO-3fb} 
of AMD calculations with constraint and superposition 
are shown by triangles with dotted line.
}
\end{figure}

In Fig. \ref{fig:be}, we show the binding energies of 
$Z=N$ nuclei from $^4$He to $^{40}$Ca obtained by 
using the present interaction parameters.
The calculations are the simple AMD calculations without constraints.
The binding energies calculated by variation after parity 
projection are shown in Fig. \ref{fig:be}(a), and the energies with
total-angular-momentum projection after the variation with parity projection 
are plotted in Fig. \ref{fig:be}(b).
We also show the results of Ref.\cite{ENYO-3fb} for the interaction(ii),
where AMD calculations with constraint and superposition were performed.
The energies for such deformed nuclei as
$^8$Be, $^{12}$C, $^{20}$Ne and $^{24}$Mg are gained by 
the total-angular-momentum projection.
As shown in Fig. \ref{fig:be}(b), the binding energies of medium $sd$ 
shell nuclei are overestimated by the interaction(i) due to the 
stronger spin-orbit force than that of the case(ii).

\subsection{Shape coexistence}
We apply the constraint AMD method to $^{40}$Ca. The width 
parameter $\nu$ for the single-particle wave packets is chosen to be 
as $\nu=0.14$ fm$^{-2}$ which optimizes the energy of the ground state.
We impose the constraint of
the total number of harmonic oscillator quanta as
$\langle \hat N_{\rm os}\rangle = N_{\rm os}$ on the parity projected 
AMD wave function, and obtain energy curves as a function of $N_{\rm os}$.
With a given constraint value $N_{\rm os}$, we obtain a few states
with different shapes as local minima.
A prolate shape and an oblate shape are obtained by randomly
choosing initial states in the variational procedure.
In order to find possible cluster states, we also start the variation
from initial states with various di-nuclear structure
and decrease the $N_{\rm os}$.
Then we find that a $\alpha$+$^{36}$Ar-like clustering appears 
as a local minimum solution.

\begin{figure}
\noindent
\epsfysize=10cm
\centerline{\epsffile{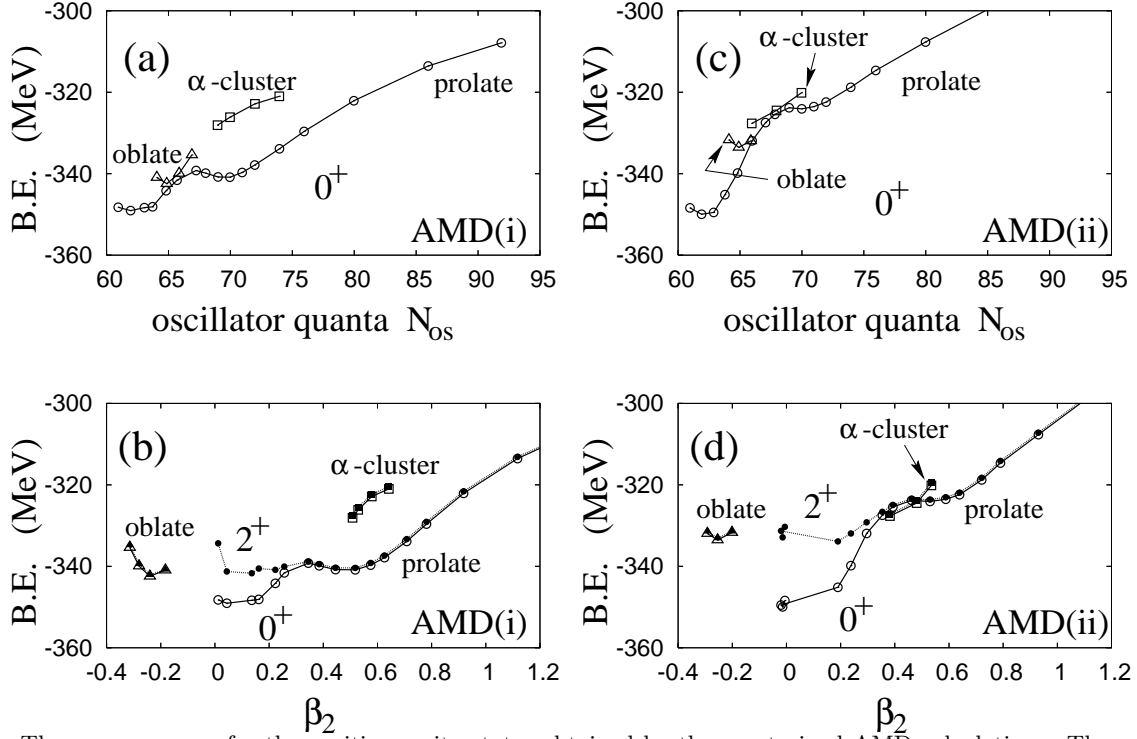}}
\caption{\label{fig:beqc-pp}
The energy curves for the positive-parity states 
obtained by the constrained AMD calculations. 
The constraint of the total number of harmonic oscillator quanta,
$\langle \hat N_{\rm os}\rangle = N_{\rm os}$, is imposed 
on the parity projected AMD wave function. 
The energies of the
spin-parity projected states, $P^{J=0(2)}_{M0}\Phi^+_{\rm AMD}
(N_{\rm os}^{(+)})$, are plotted (a)(c) as a function of $N_{\rm os}$ and
(b)(d) as a function of deformation $\beta_2$ of the states:
$\Phi^+_{\rm AMD}(N_{\rm os}^{(+)})$. The left panels((a) and (b))
show the results with the interaction(i), and the right panels((c) and (d))
are for the results with the interaction(ii).}
\end{figure}

\begin{figure}
\noindent
\epsfysize=10cm
\centerline{\epsffile{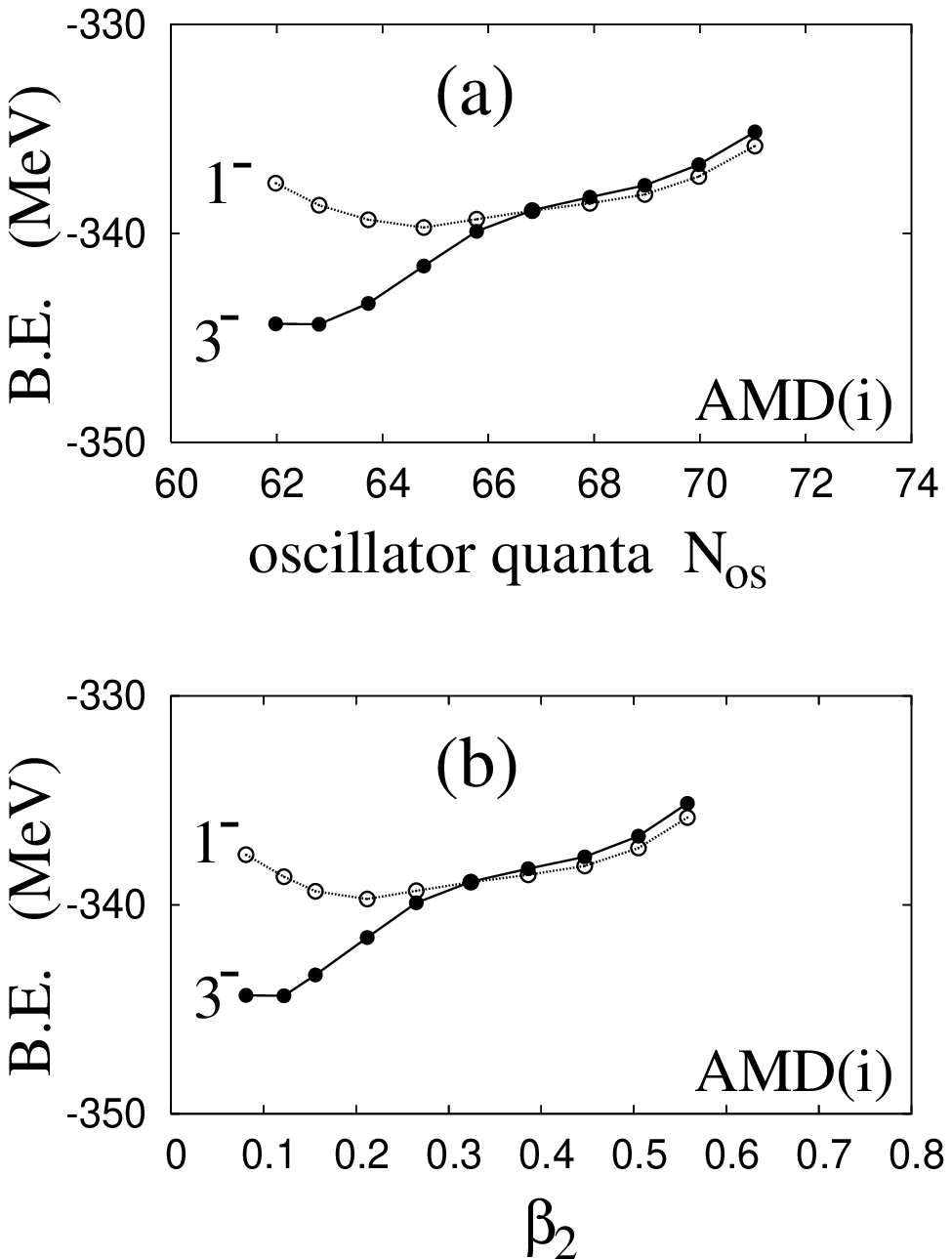}}
\caption{\label{fig:beqc-np}
The energy curves for the negative-parity states 
obtained by the constrained AMD calculations. 
The constraint of the total number of harmonic oscillator quanta,
$\langle \hat N_{\rm os}\rangle = N_{\rm os}$, is imposed 
on the parity projected AMD wave function. 
The energies of the
spin-parity projected states, $P^{J=1,3}_{M0}\Phi^-_{\rm AMD}
(N_{\rm os}^{(-)})$, are plotted as a function of $N_{\rm os}$ in the upper
panel(a) and as a function of the deformation $\beta_2$ of 
$\Phi^-_{\rm AMD}(N_{\rm os}^{(-)})$ in the lower panel(b). 
The results are obtained 
with the interaction(i).}
\end{figure}

The energy curves as a function of the oscillator quanta($N_{\rm os}$) 
are shown in Figs.\ref{fig:beqc-pp} and \ref{fig:beqc-np}.
In principle, the minimum value of $N_{\rm os}$ for 
the positive-parity state is 60, which is 
given by the $sd$-shell-closed configuration, 
and it is 61 for negative-parity states.
The results shown in the figure are calculated with the total-angular-momentum
projected wave functions
$P^J_{MK}\Phi^\pm_{\rm AMD}(N_{\rm os}^{(\pm)})$, where $K=0$ is chosen to
obtain the lowest energy.
The energies are also plotted as a function of quadrupole 
deformation $\beta_2$. Here the deformation parameter $\beta_2$ is
defined by using the sharp edge liquid drop relation between $Q_2$ 
and $\beta_2$ as follows;
\begin{eqnarray}
& \beta_2=\sqrt{\frac{5}{16\pi}}\frac{4\pi Q_2}{3 R^2 A}\\
& Q_2=2\langle z^2 \rangle-\langle x^2 \rangle- \langle y^2 \rangle,
\end{eqnarray}
where the expectation values are calculated for the intrinsic states
$\Phi^\pm_{\rm AMD}(N_{\rm os}^{(\pm)})$, and $z$ is chosen to be an
approximately symmetric axis. Here we use the fixed radius parameter, $R$, 
which is related to the nuclear mass
$A=40$ according to the formula $R=1.2 A^{1/3}$ fm.

We first mention the results of positive-parity states with the
interaction(i) (Figs. \ref{fig:beqc-pp}(a) and (b)). 
We find three states with different shapes as the local minima on the energy
surface.
The lowest state in the $J^\pi=0^+$ curve 
is an almost spherical state at $N_{\rm os}=62$. 
There are two local minimum solutions in the energy curves, 
which almost degenerate to each other, 
at 7$\sim$8 MeV higher energy
than the lowest solution.
One is an oblate state at $N_{\rm os}\sim 65$ and 
the other is a large prolate deformation at $N_{\rm os}\sim 70$ 
with $\beta_2 \sim 0.52$.
As shown later, the former contains a dominant $4p$-$4h$ configuration, 
and the latter is dominated by a `$8p$-$8h$' configuration
and corresponds to the superdeformed state. The appearance of the oblate state 
is consistent with the macroscopic-microscopic 
calculation\cite{Leander75} and also the 3d-HF 
calculations\cite{Inakura02}. On the other hand, the calculated 
energy curve for the prolate states differs from 
the macroscopic-microscopic calculation 
by Leander and Larsson\cite{Leander75} where there is no
local minimum for the $8p$-$8h$ configuration but
is a highly deformed 
$12p$-$12h$ state.
The $\alpha$+$^{36}$Ar-like cluster state appears in
$N_{\rm os}\ge 69$ region, but it is much higher than the 
superdeformed state in the case of interaction(i).
With a smaller $N_{\rm os}$ value of the constraint, the $\alpha$-cluster
is absorbed into the $^{36}$Ar core and the $\alpha$+$^{36}$Ar-like state
changes into the normal 
prolate state
during the energy variation. 
In the $2^+$ states, the energy curve for the prolate states 
is more gentle than the $0^+$ curve at the small deformation region,
$0< \beta_2 < 0.2$.

Next we look into the results obtained with
the interaction(ii) (Figs. \ref{fig:beqc-pp}(c) and (d)), which has the
weaker spin-orbit force than the interaction(i).
Although there exist the oblate state and the superdeformed state as
local minima as same as in the results(i), 
the excitation energies of these
states are much higher
than those in the case of the interaction(i).
The excitation energy of the superdeformed state is about 20 MeV, 
which largely overestimates the experimental energy, 5.21 MeV, 
of the bandhead of the superdeformed band.
On the other hand, the energy of 
$\alpha$-cluster state does not so much change from 
(i) to (ii) because the effect of the spin-orbit force is smaller than 
other excited states. As a result, $\alpha$-cluster state almost
degenerates with the superdeformed state and does not vanishes  
even in the smaller $N_{\rm os}$ region(as $N_{\rm os}\ge 66$) 
in the results(ii) than 
the case(i). Comparing Figs. \ref{fig:beqc-pp}(c) and (d), 
we find that the total number of H.O. quanta has a good 
correspondence with the quadrupole deformation $\beta_2$
in the $N_{\rm os}\ge 64$ region for the prolate deformation.
On the other hand, in the small $N_{\rm os}$ region, 
all the solutions for $N_{\rm os}\le 63$ have almost spherical shapes
with $\beta_2\sim 0$. It means that different configurations with the same
$\beta_2$ value are obtained as optimum solutions for the given $N_{\rm os}$.

\begin{figure}
\noindent
\epsfxsize=0.5\textwidth
\centerline{\epsffile{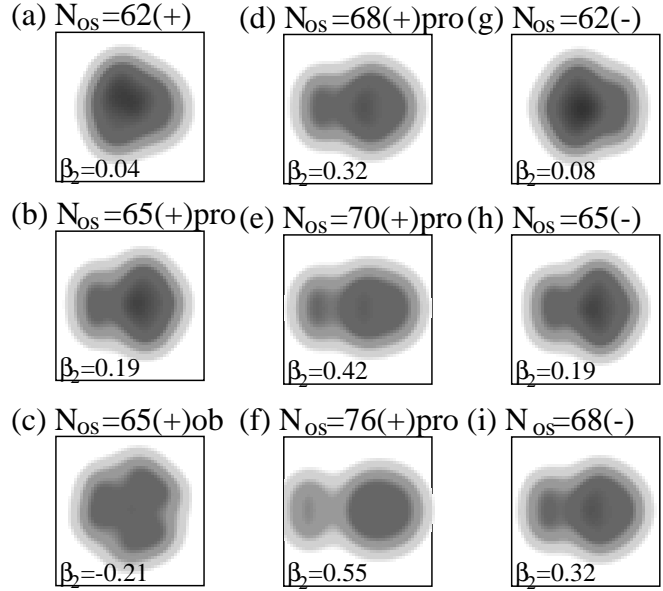}}
\caption{\label{fig:ca-dense}
Density distributions of the intrinsic states,
$\Phi_{\rm AMD}(N_{\rm os}^{(\pm)})$ for (a) the spherical state with 
$N_{\rm os}^{(\pm)}=62^{(+)}$, and the prolate solutions with
$N_{\rm os}^{(\pm)}=$
(b) $65^{(+)}$, (d) $68^{(+)}$, (e) $70^{(+)}$, (f)
 $76^{(+)}$, (g) $62^{(-)}$, (h) $65^{(-)}$, 
(i) $68^{(-)}$, and (c) the oblate solution with 
$N_{\rm os}^{(\pm)}=65^{(+)}$, which are obtained 
by using the interaction(i).
The intrinsic system is projected 
onto the $Y$-$Z$ plane, and the density is integrated along the $X$ axis,
where $X$, $Y$ and $Z$ axes are chosen as 
$\langle X^2\rangle\le \langle Y^2\rangle\le \langle Z^2\rangle$
and $\langle XY\rangle= \langle YZ\rangle= \langle ZX\rangle=0$.
The deformation parameters $\beta_2$ are written at the bottom of 
the figures. The size of the frame box is 10 fm$\times$10 fm. 
}
\end{figure}

\begin{figure}
\noindent
\epsfxsize=0.5\textwidth
\centerline{\epsffile{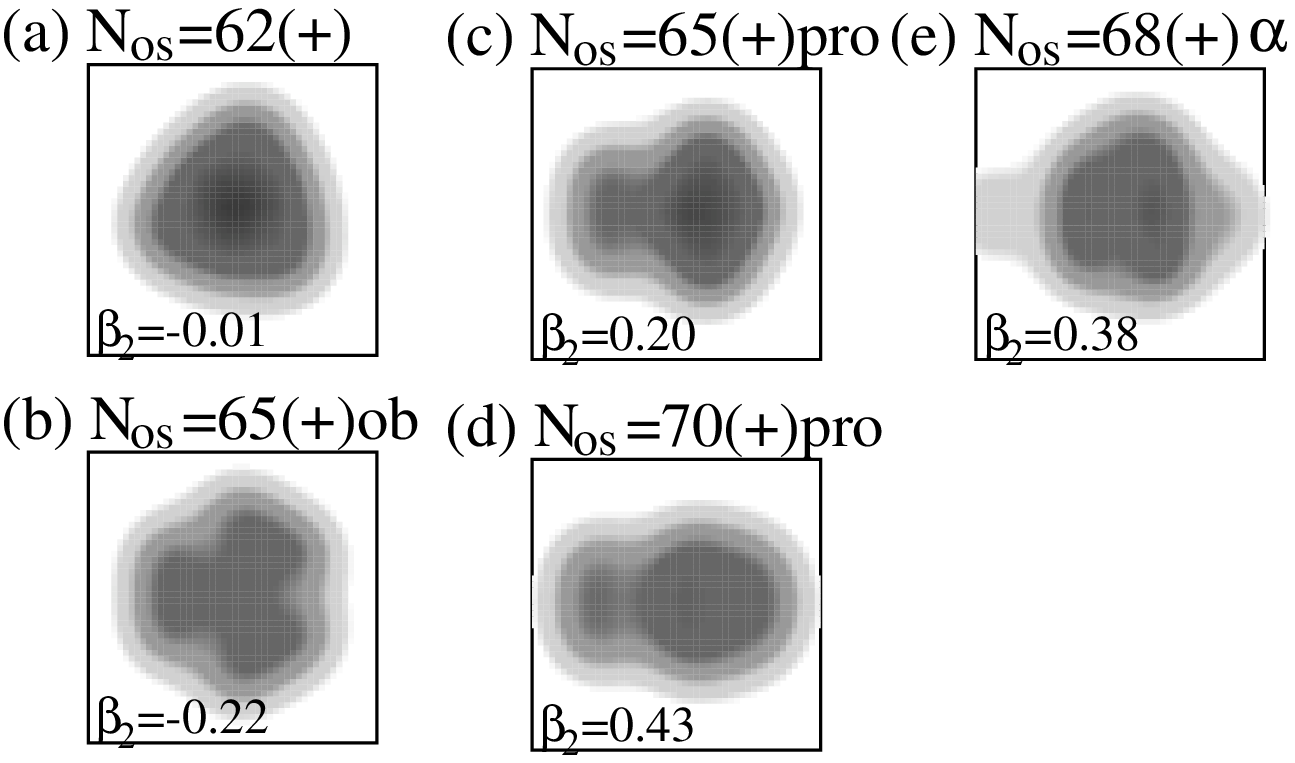}}
\caption{\label{fig:ca-dense18}
Density distribution of the intrinsic states,
$\Phi_{\rm AMD}(N_{\rm os}^{(\pm)})$ for (a) the 
the spherical state with $N_{\rm os}^{(\pm)}=62^{(+)}$,
the prolate solutions with $N_{\rm os}^{(\pm)}=$
(c) $65^{(+)}$ and (d) $70{(+)}$, (b) the oblate solution with 
$N_{\rm os}^{(\pm)}=65^{(+)}$, and (e) the $\alpha$-cluster state
with $N_{\rm os}^{(\pm)}=68^{(+)}$, which are
 obtained by using the interaction(ii).
The intrinsic system is projected 
onto the $Y$-$Z$ plane, and the density is integrated along the $X$ axis,
where $X$, $Y$ and $Z$ axes are chosen as 
$\langle X^2\rangle\le \langle Y^2\rangle\le \langle Z^2\rangle$
and $\langle XY\rangle= \langle YZ\rangle= \langle ZX\rangle=0$.
The deformation parameters $\beta_2$ are written at the bottom of 
the figures. The size of the frame box is 10 fm$\times$10 fm. 
}
\end{figure}

We illustrate the density distribution of the intrinsic state 
$\Phi_{\rm AMD}(N_{\rm os}^{(\pm)})$ 
in Fig.\ref{fig:ca-dense} and Fig.\ref{fig:ca-dense18}.
A remarkable point is that parity asymmetric structure appears in the
prolate states. 
Especially, one of striking features is the prominent 
cluster structure like $^{12}$C+$^{28}$Si in the large prolate deformation.
Comparing the results of (i) and (ii), 
we remark that the intrinsic structure for a given $N_{\rm os}$
is almost the same as each other. 
It means that the structure of coexisting shapes is not 
sensitive to the choice of either interaction 
(i) or (ii) except for the $\alpha$-cluster state.

In contrast, the excitation energies strongly depend 
on the strength of the spin-orbit force
as shown in the comparison between the results 
with a stronger spin-orbit force(i) and with a weaker one(ii). 
Particularly, the oblate state 
and the superdeformed state appear at the low excitation energies in 
the results with the interaction(i), but they exist
at much higher excitation energies with the interaction(ii).
Considering the $4p$-$4h$ configuration in the oblate shape and
the $^{12}$C-$^{28}$Si clustering in the superdeformed state, it is
easily understood that the excitation energies of these states 
decrease with the stronger spin-orbit force(i) than with the weaker one(ii).
Although the systematics of the binding energy in the $sd$-shell region
seems to be reproduced better by the force(ii) than the force(i), however,
the excitation energy of the superdeformed state
is reasonably reproduced with 
the force(i) while they are largely overestimated by the force(ii).
The force(i) also reproduces the excitation energy of the lowest
negative-parity state($J^\pi=3^-$) rather well than the force(ii)
as shown later. 
Therefore, we examine mainly the results(i), 
because our interest is in excited states of $^{40}$Ca, 
especially in the properties of the superdeformed state.
We should note again that 
the structure of the superdeformed state is stable in both cases 
(i) and (ii), and 
most of the intra-band properties of the superdeformed band does not so much 
depend on the choice of the interaction(i) or (ii) except for 
the relative energy to the ground state and other rotational bands.

Now we turn to negative-parity states.
In Fig.\ref{fig:beqc-np}, we show the energies
of $J^\pi=1^-$($P^{J=1}_{MK}\Phi^-_{\rm AMD}(N_{\rm os}^{(-)})$) and 
$3^-$($P^{J=3}_{MK}\Phi^-_{\rm AMD}(N_{\rm os}^{(-)})$),
obtained with the interaction(i),
as functions of (a) $N_{\rm os}$ and (b) $\beta_2$.
It is found that the constraint on $\langle \hat N_{\rm os}\rangle$ acts 
as the constraint on deformation $\beta_2$ 
in negative-parity states as well as in the positive-parity ones.
In the small deformation region, the lowest state is 
the $J^\pi=3^-$ which corresponds to a $1p$-$1h$ state.
The difference between the minimum energy of the $3^-$ state and that of 
the $0^+$ state is 4.7 MeV, 
which well corresponds to the 
experimental excitation energy 3.74 MeV of the lowest $3^-$ state.
We here comment that the interaction (ii) gives a much larger 
energy difference as 6.9 MeV than the interaction (i) and fails to reproduce
the experimental excitation energy of the $3^-$ state.
As $N_{\rm os}$ becomes large, the $J^\pi=1^-$ state becomes lower than
the $J^\pi=3^-$ state and a $K^\pi=0^-$ rotational band with a 
prolate intrinsic state is built on this state.
We illustrate the density distributions of the intrinsic states for 
the negative parity states, $\Phi_{\rm AMD}(N_{\rm os}^{(-)})$ 
in Fig.\ref{fig:ca-dense}(g)-(i).
For each $N_{\rm os}$, it is found that intrinsic structure 
of $\Phi_{\rm AMD}(N_{\rm os}^{(-)})$ for
the negative-parity state is quite similar to that of the prolate solution 
$\Phi_{\rm AMD}(N_{\rm os}^{(+)})$ for the positive-parity state.
In particular, the negative-parity state with $N_{\rm os}\ge 68$ has
the parity-asymmetric shape because of the clustering which is regarded as
the $^{12}$C+$^{28}$Si-like structure as well as the case of positive parity 
state. 
As a result, the largely deformed state forms the $K^\pi=0^-$ rotational band 
consisting of $J^\pi=1^-, 3^-, \cdots$  
due to the $Y_{30}$ deformation, which may have a link to
the positive-parity bands as a parity doublet.

\subsection{Energy levels}

In the obtained wave function $\Phi_{\rm AMD}(N_{\rm os}^{(\pm)})$,
there exist local minimum solutions in the energy curve as a function
of $N_{\rm os}$. They are also local minima 
with respect to the deformation $\beta_2$ and considered to
be approximate intrinsic wave functions for the corresponding 
deformed bands. In order to satisfy the orthogonality among the levels
and to obtain better wave functions, 
we calculate the level scheme by superposing 
the spin-parity projected wave functions,
$P^J_{MK}\Phi^\pm_{\rm AMD}(k)$.

\begin{figure}
\noindent
\epsfxsize=0.4\textwidth
\centerline{\epsffile{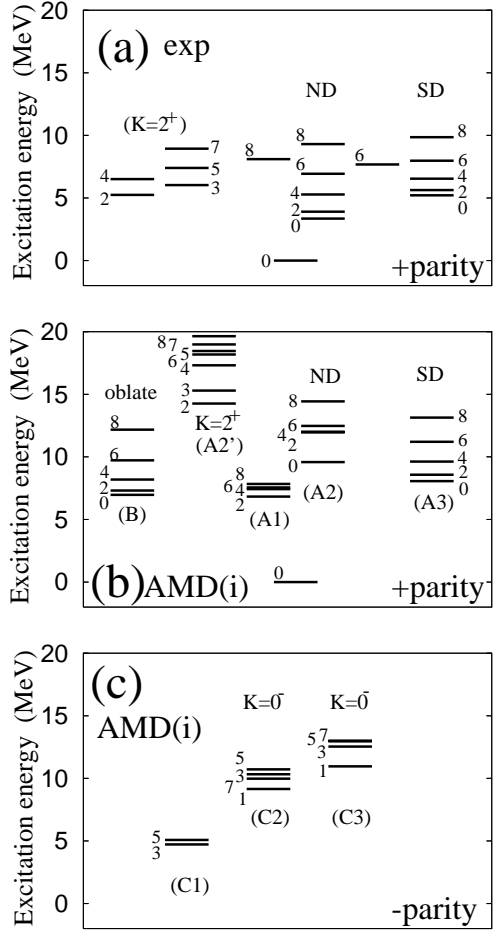}}
\caption{\label{fig:level}
Excitation energies of the levels up to $J=8$ in $^{40}$Ca.
The experimental data of the levels observed in 
Ref.\protect\cite{Ideguchi01} are shown in (a). Figures (b) and (c) show
the theoretical results for the positive-parity and negative-parity 
states, respectively, calculated by the superposition 
of the AMD wave functions.
The interaction(i) is used. The excitation energies of the 
higher spin($J>8$) states will be shown in the next figures.}
\end{figure}

In the calculations with the case(i) interaction, 
we use $N_{\rm os}^{(\pm)}=$$62^{(+)}$, 
$65^{(+)}$, $68^{(+)}$, $70^{(+)}$, $72^{(+)}$, $76^{(+)}$, 
$80^{(+)}$, $86^{(+)}$, $92^{(+)}$ with prolate shapes  and
$N_{\rm os}^{(\pm)}=65^{(+)}_{\rm ob}$ with the oblate shape 
as the basis for the positive-parity states. 
For the negative-parity states, we adopt 
$N_{\rm os}^{(\pm)}=$$62^{(-)}$, $63^{(-)}$, $65^{(-)}$, $66^{(-)}$,
$68^{(-)}$, $69^{(-)}$, $71^{(-)}$, $65^{(+)}$, $68^{(+)}$, 
$65^{(+)}_{\rm ob}$.
The obtained level scheme is shown in Fig.\ref{fig:level}.
By analysing such properties as the dominant components 
and $E2$ transition strengths of the obtained levels,
we classify the levels into groups, which are labeled in 
Fig.\ref{fig:level}(b) and (c).

\begin{figure}
\noindent
\epsfxsize=0.4\textwidth
\centerline{\epsffile{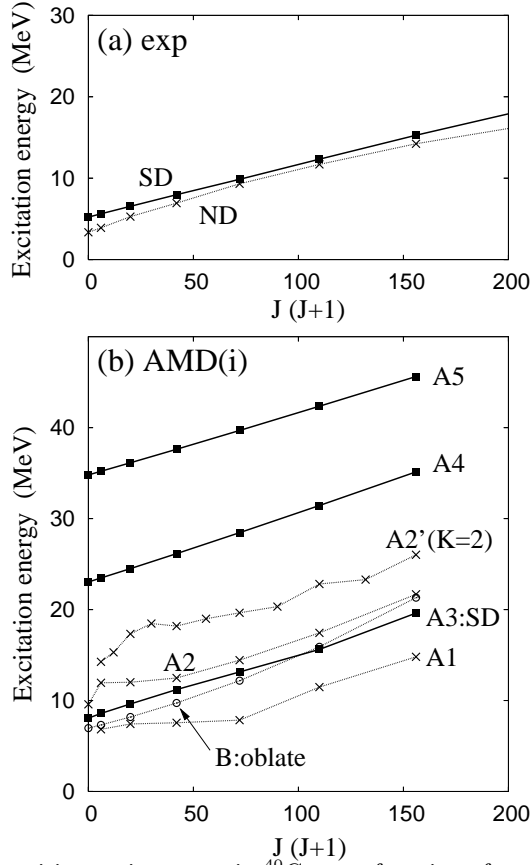}}
\caption{\label{fig:band}
Excitation energies of the positive-parity states in $^{40}$Ca 
as a function of a spin $J(J+1)$. Theoretical values shown 
in the lower panel (b) are those obtained with the interaction(i).
The energies are calculated by the superposition 
of the AMD wave functions.
The upper panel (a) shows the experimental data for the normal-deformed band
and the superdeformed band, which are labeled ``band 2'' and ``band 1''
in Ref.\protect\cite{Ideguchi01}, respectively.}
\end{figure}

In the positive-parity levels,
we find $K^\pi=0^+$ rotational bands  
corresponding to the shape coexistence:(B)the oblate band
and (A3)the superdeformed band. The states in the groups (A1) and (A2)
are built from the mixing of the spherical intrinsic state 
$\Phi_{\rm AMD}(N_{\rm os}=62^{(+)})$ and the 
normal prolate state $\Phi_{\rm AMD}(N_{\rm os}=65^{(+)})$, while
the side band (A2'(K=2)) of the normal deformation appears 
from the dominant $P^J_{M2}\Phi^+_{\rm AMD}(N_{\rm os}=65^{(+)})$ component
due to the triaxial nature of the intrinsic state. 
In Fig.\ref{fig:band}, we plot 
the calculated excitation energies as a function of a spin $J(J+1)$.
We also show 
the experimental data of the bands labeled ``band 2'' and ``band 1'' 
in Ref.\cite{Ideguchi01}, which are assigned to be 
the normal-deformed band and the superdeformed band, respectively.
Since the band (A2)
contains the configuration mixing, the calculated level structure 
is not consistent
with the rigid rotor levels and does not well agree
to the experimental data of the normal-deformed band 
(labeled ``band 2'' in Ref.\cite{Ideguchi01}). 
Moreover, it is higher than the superdeformed band in the present results(i),
however, we tentatively assign this band as the normal-deformed band 
because of rather strong $E2$ transitions in (A2) band.
On the other hand, the calculated energy spectra of the 
superdeformed band behave as that of the rigid rotor and 
agrees to the experimental data for ``band 1'', 
though the excitation energy of the band-head
state is slightly higher than the experimental data.
The dominant components of the superdeformed band 
in the low-spin levels 
is $P^J_{M0}\Phi^+_{\rm AMD}(N_{\rm os}=70^{(+)})$, which have a 
$^{12}$C+$^{28}$Si-like clustering. 
Another interesting character of the superdeformed band is its particle-hole
property. As discussed later in detail, 
$P^J_{M0}\Phi^+_{\rm AMD}(N_{\rm os}=70^{(+)})$
has about 90\% overlap 
with the band-head of the superdeformed band, and its intrinsic state  
$\Phi_{\rm AMD}(N_{\rm os}=70^{(+)})$ is regarded as a
``$8p$-$8h$'' state. This particle-hole feature
of the superdeformed band is supported by the experimental fact that
the $0^+($5.21 MeV), $2^+($5.63 MeV), and $4^+($6.54 MeV) in the 
superdeformed band are strongly populated 
in 8-nucleon-transfer reactions, 
$^{32}$S($^{12}$C,$\alpha$)$^{40}$Ca\cite{Middleton72}.
However, it is not consistent with 
the GCM calculations of Skyrme HF+BCS\cite{Bender03},
where the superdeformed band can not be described 
by a certain $np$-$nh$ configuration 
but contains a mixture of various configurations such as
$4p$-$4h$, $6p$-$6h$, and $8p$-$8h$ states.
In order to check the stability of the superdeformed band for the choice of
the intrinsic wave functions to be superposed, 
we use another set of the basis as 
$\{k\}=\{ N_{\rm os}^{(\pm)}\}=\{$$62^{(+)}$, $64^{(+)}$, $65^{(+)}$, 
$66^{(+)}$, $67^{(+)}$, $68^{(+)}$, $70^{(+)}$, $72^{(+)}\}$ and find that
the level scheme and the $E2$ transition strengths 
in the superdeformed band are stable against the choice of the basis,
and the properties of the states in bands (A1) and (A2) are qualitatively
unchanged. In the presetn calculations, the number of the basis
in the superposition is 10 at most and is much smaller 
than the GCM calculations in Ref.\cite{Bender03}. 
It also should be noted that, we did not obtain the basis 
with the quadrupole deformation $0 <\beta_2< 0.2$ as shown in
Fig.\ref{fig:beqc-pp}. For the detailed investigation of 
the band mixing, it is expected to be important to perform 
GCM calculations with a large number of the basis concerning 
the two-dimensional constraints on the $N_{\rm os}$ and $\beta_2$.
The present results of the weak mixings 
between different $np-nh$ configurations 
seem to be consistent with the experimental facts, however, 
we comment that the pairing correlations, which were taken into account
in Ref.\cite{Bender03}, are ignored 
in the present calculations though we think they are partially 
included through the spin-parity projections and superposition 
of the basis. Since the pairing correlations increase configuration
mixings in general, it would be valuable to estimate 
the pairing effects on the stability of the superdeformed band.

In addition to the prolate bands,
the oblate band appears at almost the same excitation energy 
as that of the superdeformed band. The moment of inertia is smaller 
than the superdeformed band. The oblate deformation is predicted also 
in the macroscopic-microscopic calculations\cite{Leander75} and
in the mean-field calculations\cite{Bender03,Inakura02}. However,  
since an oblate band has not been experimentally assigned yet,
the existence of an oblate shape in $^{40}$Ca is an open problem. 
We also find highly excited bands (A4) and (A5) which arise from 
the developed $^{12}$C+$^{28}$Si clustering. The details of these bands 
are discussed in the next section.

In the negative-parity states(Fig.\ref{fig:level}c),
the lowest $3^-$ and $5^-$ states appear from almost the spherical states.
The theoretical excitation energies are in good agreement with the
experimental data of the $3^-_1$(3.74 MeV) and 
the $5^-_1$(4.49 MeV).
We find two $K^\pi=0^-$ rotational bands (C2) and (C3) which arise 
from the parity-asymmetric shape in the prolate states.
Since the energies of the $3^-$ and $5^-$ states in these (C2) and (C3) 
bands are raised by the mixing with the lowest $3^-$ and $5^-$ states, 
the level structure is somehow out of that for the typical $K^\pi=0^-$ 
rotational band.
The $1_1^-$ state($\Psi_{n=1}^{1-}$) in the band (C2) is dominated by 
$P^{J=1}_{M0}\Phi^-_{\rm AMD}(N_{\rm os}=65^{(-)})$ 
with the overlap 
$|\langle \Psi_{n=1}^{1-}|P^{J=1}_{M0}\Phi^-_{\rm AMD}(N_{\rm os}=65^{(-)}) 
 \rangle |^2\sim$ 90\%, while the $1^-_2$ state($\Phi_{n=2}^{1-}$) 
in the band (C3) has the dominant
$P^{J=1}_{M0}\Phi^-_{\rm AMD}(N_{\rm os}=69^{(-)})$ 
with 70\% overlap. We should note that the $1_1^-$ and $1^-_2$ 
states have large overlap with also the $1^-$ states 
projected from the intrinsic states for the positive-parity,
$\Phi_{\rm AMD}(N_{\rm os}=65^{(+)})$ and 
$\Phi_{\rm AMD}(N_{\rm os}=68^{(+)})$, respectively.
It is reasonable because 
$\Phi_{\rm AMD}(N_{\rm os}^{(-)})$ is almost the same state as the 
$\Phi_{\rm AMD}(N_{\rm os}^{(+)})$ when the same constraint 
$\langle \hat N_{\rm os}\rangle=N_{\rm os}$ is imposed. 
In order to find the possible parity doublet of the superdeformed band,
we pay special attention to the 
$\Phi_{\rm AMD}(N_{\rm os}=68^{(+)})$ component in the 
largely deformed negative-parity band (C3) and in 
the superdeformed band (A3) with positive-parity.
The overlap $|\langle \Phi_{n=2}^{1-}|P^{J=1}_{M0}
\Phi^-_{\rm AMD}(N_{\rm os}=68^{(+)}) 
 \rangle |^2$ in the bandhead of the $K^\pi=0^-$ band(C3) 
is about 70\%, while the bandhead $0^+$ of the superdeformed 
band(A3) has 80\% overlap with 
$P^{J=0}_{M0}\Phi^+_{\rm AMD}(N_{\rm os}=68^{(+)})$.
Then it leads to an interpretation that the largely deformed negative-parity 
band (C3) and the positive-parity superdeformed band (A3) 
are approximately the parity doublets $K^\pi=0^-$ and $K^\pi=0^+$
arising from the intrinsic state $\Phi_{\rm AMD}(N_{\rm os}=68^{(+)})$, 
which has the parity asymmetric shape with the 
$Y_{30}$ deformation due to the $^{12}$C+$^{28}$Si-like clustering
as shown in Fig.\ref{fig:ca-dense}. 

\begin{figure}
\noindent
\epsfxsize=0.4\textwidth
\centerline{\epsffile{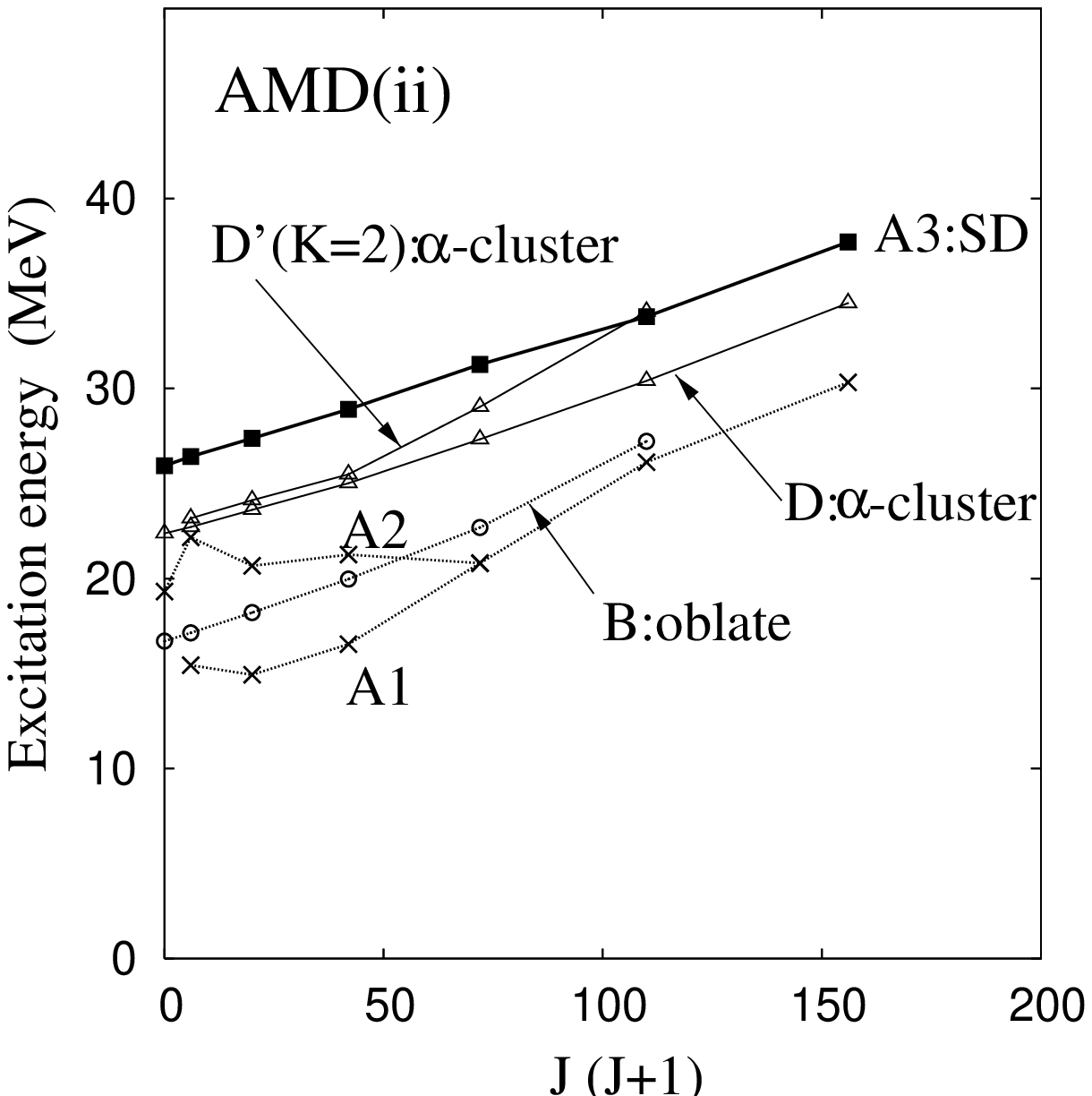}}
\caption{\label{fig:band-2}
Excitation energies of the positive-parity states in $^{40}$Ca 
obtained with the interaction(ii).
The energies are calculated by the superposition 
of the AMD wave functions.}
\end{figure}

Next we discuss the energy levels calculated with the interaction(ii).
We use $N_{\rm os}^{(\pm)}=$$62^{(+)}$,
$65^{(+)}$, $68^{(+)}$, $70^{(+)}$, $72^{(+)}$ with prolate deformations, 
$N_{\rm os}^{(\pm)}=65^{(+)}_{\rm ob}$ with the oblate shape, and
$N_{\rm os}^{(\pm)}=66^{(+)}_\alpha$, $68^{(+)}_\alpha$, $70^{(+)}_\alpha$ with $\alpha$-cluster-like
structure 
as the basis in superposing $P^{J}_{MK}\Phi^+_{\rm AMD}(k)$.
The calculated excitation energies are shown in Fig.\ref{fig:band-2}. 
The rotational bands 
of oblate deformation, prolate one and superdeformation coexist 
as well as in the results(i).
The excitation energies of these bands with the interaction(ii) are 
much higher than those of the results(i), and also are inconsistent with
those of the experimentally known low-lying states.
For example, the oblate band (B) starts from 
17 MeV excitation energy, 
while the bandhead energy of the superdeformed band (A3) is 
26 MeV.
One of the characteristics of the results(ii) is that 
a $\alpha$-cluster-like band (D) appears at the 
relatively lower energy than the superdeformed band.

It has been known that the low-spin states in the normal-deformed 
band built on the $0^+_2$(3.35 MeV) state are strongly populated 
in $\alpha$-transfer reactions, and they have relatively
larger $\alpha$-spectroscopic factors than other low-lying states
\cite{Betts77,Fortune79,Yamaya94,Yamaya98}. 
Therefore, it is expected that the components of 
the $\alpha$-cluster state should be contained in the 
normal-deformed band. However, in the present calculations 
we failed to describe the $\alpha$-cluster
components in the low-lying states, and
even in the results(ii), the components of the $\alpha$-cluster state 
do not mix with the normal-deformed band (A2) except for the $2^+$ state.
It is because the energy difference between the bands (A2) and (D) are large
in the present calculations. In order to solve this problem, it may be 
necessary to introduce a suitable nuclear interaction and extended model 
wave functions.

\subsection{$E2$ transitions and moments}

We calculate the $E2$ transition strengths and the electric 
quadrupole moments for the excited states.
We also extract quantities related to the intrinsic deformation 
from the physical observables such as $B(E2)$ and electric quadrupole moments. 
By using the collective model approximation
\cite{bohr}, we define the intrinsic quadrupole moment and the deformation
by the $E2$ transitions as, 
\begin{equation}\label{eq:Q2t}
Q^{t}_0=\sqrt{\frac{16\pi}{5e^2}\frac{B(E2,J_i\rightarrow J_f)}
{\langle J_i K 2 0 |J_f K\rangle ^2}},\quad 
\beta^{t}=\sqrt{\frac{5}{16\pi}}\frac{4\pi Q^{t}_0}{3 R^2 Z},
\end{equation}
where $\langle J_i K 2 0 |J_f K\rangle$ is a Clebsch-Gordan coefficient.
$t$ stands for ``transition''. 
We also extract the intrinsic quantities from the observable 
electric quadrupole moments $Q$ as follows,
\begin{equation}
Q^{s}_0=\frac{1}{e}\frac{(J+1)(2J+3)}{3K^2-J(J+1)} Q,
\quad\beta^{s}=\sqrt{\frac{5}{16\pi}}\frac{4\pi Q^{s}_0}{3 R^2 Z}.
\end{equation}
Here $s$ stands for ``spectroscopic''.

In the table \ref{tab:be2}, \ref{tab:be2np} and \ref{tab:be2-18},
we show the calculated $B(E2)$ and $Q$ moments, and related quantities.
The $E2$ transitions are small among the states in the group (A1), which
are dominated by spherical states and considered not to be a 
rotational band of a deformed state.
On the other hand, the prolate band (A2) and the oblate one(B) have moderate
transition strengths, while the $E2$ transitions 
are remarkably strong in the superdeformed band (A3). 

\begin{figure}
\noindent
\epsfxsize=0.4\textwidth
\centerline{\epsffile{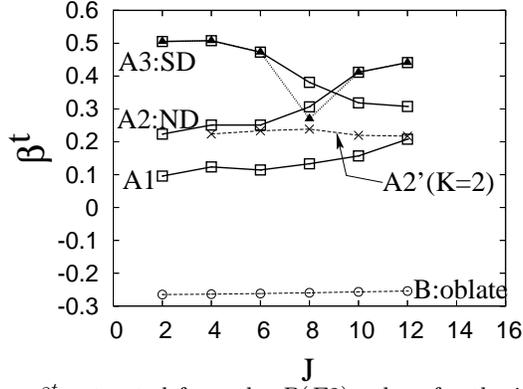}}
\caption{\label{fig:beta}
The deformation parameters $\beta^t$ extracted from the $B(E2)$ values
for the intra-band transitions.
The labels of the bands correspond to those in 
Fig.\protect\ref{fig:level}.
The results are for the calculations with the interaction(i).
}
\end{figure}

In the results with the interaction(i),
the deformation parameter $\beta^t$ extracted from the $B(E2)$ is
consistent with $\beta^s$ evaluated with the $Q$ moment 
except for negative-parity bands and high spin states in the prolate
bands (A1),(A2), and (A3). 
In Fig.\ref{fig:beta}, we show the behavior of the transition deformation 
$\beta^t$ with the increase of 
the total-angular momentum $J$. 
In the oblate band (B), the intrinsic 
deformation $\beta$ is almost constant with the increase of $J$,
because the oblate states do not mix with the other shapes.
On the other hand, $\beta^t$ in the 
bands (A1), (A2) and (A3)
changes as $J$ increases.
The deformation of the low-spin states in the superdeformed band is
the largest as $\beta^t\approx 0.5$.
With the increase of $J$,
$\beta^t$ of the superdeformed band (A3) decreases, 
while that of the normal-deformed band (A2) enlarges, 
and the $\beta^t$-lines for these two bands 
cross each other around $J=8$. 
We remark that the state mixing occurs in 
the region $J\ge 6$, and the 
components contained in (A2) and (A3) change around $J=8$.
As a result, the transition from the $8^+(A2)$ to $6^+(A3)$ is 
significant strong(see Table \ref{tab:be2}). 
Therefore, another assignment may be possible 
as $8^+(A2)$, $10^+(A2)$ and $12^+(A2)$ belong to the superdeformed band.
In fact, 
$\beta^t$ extracted from the $B(E2;J\rightarrow J-2)$ among the 
$0^+(A3)$, $2^+(A3)$, $4^+(A3)$, $6^+(A3)$, $8^+(A2)$, $10^+(A2)$ 
and $12^+(A2)$ are so large as shown by filled triangles 
in Fig.\ref{fig:beta}, 
that these states can be assigned to compose a rotational band 
with the strong $E2$ transitions.
In the experimental data by Ideguchi et al.\cite{ideguchi01},
they observed the deviation from the constant 
moment of inertia in the superdeformed band 
and also the inter-band gamma transitions 
between normal deformation and superdeformation. These 
facts may give an indication of the possible band mixing or 
some structure change around $J\sim 8$ in the superdeformation.
Further fine measurements of $E2$ transition strengths will be 
helpful to see the details of the band mixing and to establish 
the assignment for the high spin states of the superdeformed band.

In the results with the interaction(ii) (table \ref{tab:be2-18}), 
we also find strong $E2$ transitions in the superdeformed band
(A3). The transitions of the prolate band (A2) are weaker in the results(ii)
than in the case(i). As mentioned before, 
the $K^\pi=0^+$(D) and $K^\pi=2^+$(D') bands with 
the $\alpha$-cluster structure appear below the superdeformed band in the 
results(ii). These bands have strong intra-band $E2$ transitions, which give
the larger transition deformations $\beta^t\approx 0.4$
than the prolate band (A2).

Let us make a comparison of intra-band transitions between the present results 
and the experimental data, and also compare them with other theoretical calculations.
The theoretical and experimental values of $Q^t_0$ for the 
normal-deformed band and the
superdeformed band are listed in table \ref{tab:Q2t}.
The present results for the superdeformed band are in reasonable
agreements with the experimental data.
They are also consistent with the results by GCM+HFBCS calculations. 
In the shell model calculations by Poves {\em et al.}, 
large $Q^t_0$ values were obtained with the truncated calculations
within $8p$-$8h$ configurations by using effective charges $q_\pi=1.5$
and $q_\nu=0.5$ for protons and neutrons, however, the $Q^t_0$ 
get eroded when other particle-hole configurations are mixed in the calculations.
The $Q^t_0$ moments of the band (A2) are $Q^t_0=57-112$ fm$^2$ and
comparable to the experimental values $Q^t_0=74\pm 14$ fm$^2$
for the normal deformation(band 2) in Ref.\cite{Ideguchi01}, but are smaller
than the experimental data $Q^t_0=1.1\pm 0.1\times 10^2$ 
and $1.3\pm0.2 times 10^2$ fm$^2$ in Ref.\cite{nucldata}.
Our results are consistent with the GCM+HFBCS calculations except for
the $4^+$ state, which has a small $Q^t_0$ in the GCM+HFBCS result.
As shown in table \ref{tab:be2-18}, we obtain 
the $\alpha$-cluster band (D) with strong $E2$ transitions in case of 
the interaction(ii). The transition moment of the band (D) is 
$Q^t_0\approx 100$ fm$^2$ and well agrees to the experimental data 
in Ref.\cite{nucldata}. Therefore, 
there might be an alternative assignment that 
the band (D) corresponds to the normal-deformed band built on the 
$0^+_2$(3.35 MeV), though the excitation energy of the band (D) 
is much higher than the experimental data in the present calculations.

The $K=2$ band were experimentally established based on the measurement of 
$B(E2)$ \cite{Macdonald,Wood92}. 
The observed intra-band transitions are 
$B(E2;3^+ \rightarrow 2^+)=730\pm 300$ ($e^2$ fm$^4$) and
$B(E2;4^+ \rightarrow 2^+)=200+150-75$ ($e^2$ fm$^4$) 
\cite{Macdonald}.
The properties of this band were
studied by Sakuda and Ohkubo with $\alpha$+$^{36}$Ar($2^+$)-cluster model
\cite{Sakuda94,Sakuda98}, which well repoduces these intra-band transitions.
In the present calculations, we find 
the $K^\pi=2^+$ band (A2') with the prolate deformation in the results(i),
and the $K^\pi=2^+$ band (D') with the $\alpha$-cluster structure in the
case of (ii). In the band (A2'), the calculated $E2$ transition strengths are
$B(E2;3^+\rightarrow 2^+$=95 ($e^2$ fm$^4$) and 
$B(E2;4^+\rightarrow 2^+$=38 ($e^2$ fm$^4$), which are smaller than 
experimental $B(E2)$ values. On the other hands, the 
$E2$ transition strengths in the band (D') are large enough as 
$B(E2;3^+ \rightarrow 2^+)=390$ ($e^2$ fm$^4$) and
$B(E2;4^+ \rightarrow 2^+)=129$ ($e^2$ fm$^4$), which well correspond
to the data within the experimental error bars. It means that
the deformation of the $K=2$ band can be described by the $\alpha$-cluster 
band (D'), which is obtained with the interaction (ii).

Next we discuss the moments of the negative-parity states.
For the negative-parity states in the group (C1),
the magnetic dipole moments of the $3^-$ and $5^-$ states are 
$\mu(3^-)=1.66\ \mu_N$ and $\mu(5^-)=2.76\ \mu_N$, which well agree
to the experimental values for the lowest $3^-_1$(3.74 MeV) and 
the lowest $5^-_1$(4.49 MeV);
$\mu(3_1^-)=1.65\ \mu_N$ and $\mu(5^-)=2.6\ \mu_N$.
In the prolate bands (C2) and (C3),
the $E2$ transitions are fragmented among the bands
because the $3^-$ and $5^-$ states of these bands 
contain configuration mixing due to
the existence of the lowest $3^-$ and $5^-$ states in the group (C1) 
(see table \ref{tab:be2np}).
Then, we estimate the intrinsic deformation 
by looking into $\beta^s$ extracted from the $Q$ moments instead of $\beta^t$.
According to the calculated $\beta^s$ of the band-head state($1^-$), 
the deformation of the band (C3) is evaluated as 
$\beta^s=0.39$. This is smaller than that of the 
superdeformed band (A3), which we raughly assign to the parity doublet of (C3).

\begin{table}
\caption{ \label{tab:be2} Electric quadrupole moments $Q$ ($e$ fm$^2$) and $B(E2)$ 
($e^2$fm$^4$) for the positive-parity states 
calculated with the interaction(i). 
The extracted intrinsic quadrupole moments (fm$^2$) and deformation
parameters are also listed. The definitions of $Q^{t,s}_0$ and $\beta^{t,s}$
are explained in the text.
The labels of the bands correspond those in 
Fig.\protect\ref{fig:level}.}
\begin{center}
\begin{tabular}{cccccccccc}
& & & & & \multicolumn {3}{l}{tansition to}& & \\
$J^\pi$ & band    & $Q$    & $Q^s_0$ & $\beta^s$ & final & band    & $B(E2)$ & $Q^t_0$ 
& $\beta^t$ \\
\hline
\multicolumn{3}{l}{positive-parity bands}   &      &      &       &         &     &     &      \\
\hline
$2^+$   & (A1)    &  $-$10 &   34 & 0.13 & $0^+$ & (A2)    &  12 &  24 & 0.10 \\
$4^+$   & (A1)    &  $-$10 &   28 & 0.11 & $2^+$ & (A1)    &  28 &  31 & 0.12 \\
$6^+$   & (A1)    &  $-$13 &   33 & 0.13 & $4^+$ & (A1)    &  27 &  29 & 0.11 \\
$8^+$   & (A1)    &  $-$15 &   36 & 0.14 & $6^+$ & (A1)    &  38 &  34 & 0.13 \\
$10^+$  & (A1)    &  $-$22 &   50 & 0.19 & $8^+$ & (A1)    &  54 &  40 & 0.16 \\
$12^+$  & (A1)    &  $-$30 &   68 & 0.27 & $10^+$& (A1)    &  96 &  53 & 0.21 \\
\hline
\multicolumn{3}{l}{prolate} &      &      &       &         &     &     &      \\
$2^+$   & (A2)    &  $-$17 &   61 & 0.24 & $0^+$ & (A2)    &  64 &  57 & 0.22 \\
$4^+$   & (A2)    &  $-$24 &   65 & 0.26 & $2^+$ & (A2)    & 116 &  64 & 0.25 \\
$6^+$   & (A2)    &  $-$29 &   72 & 0.28 & $4^+$ & (A2)    & 128 &  64 & 0.25 \\
$8^+$   & (A2)    &  $-$43 &  102 & 0.40 & $6^+$ & (A2)    & 200 &  78 & 0.31 \\
$8^+$   & (A2)    &   &  &  & $6^+$ & (A3)    & 161 &  70 & 0.27 \\
$10^+$  & (A2)    &  $-$50 &  116 & 0.45 & $8^+$ & (A2)    & 370 & 105 & 0.41 \\
$12^+$  & (A2)    &  $-$56 &  126 & 0.49 & $10^+$& (A2)    & 433 & 112 & 0.44 \\
\hline
\multicolumn{3}{l}{superdeformation}  &      &      &       &         &     &     &      \\
$2^+$   & (A3)    &  $-$37 &  131 & 0.51 & $0^+$ & (A3)    & 330 & 129 & 0.51 \\
$4^+$   & (A3)    &  $-$47 &  129 & 0.51 & $2^+$ & (A3)    & 476 & 129 & 0.51 \\
$6^+$   & (A3)    &  $-$47 &  118 & 0.46 & $4^+$ & (A3)    & 455 & 121 & 0.47 \\
$8^+$   & (A3)    &  $-$40 &   95 & 0.37 & $6^+$ & (A3)    & 309 &  97 & 0.38 \\
$10^+$  & (A3)    &  $-$36 &   83 & 0.33 & $8^+$ & (A3)    & 222 &  81 & 0.32 \\
$12^+$  & (A3)    &  $-$43 &   97 & 0.38 & $10^+$& (A3)    & 210 &  78 & 0.31 \\
\hline
\multicolumn{3}{l}{oblate}  &      &      &       &         &     &     &      \\
$2^+$   & (B)     &   19 &  $-67$ & $-$0.26& $0^+$ & (B)     &  91 &  $-68$ & $-0.26$ \\
$4^+$   & (B)     &   24 &  $-66$ & $-$0.26& $2^+$ & (B)     & 128 &  $-67$ & $-0.26$ \\
$6^+$   & (B)     &   25 &  $-63$ & $-$0.25& $4^+$ & (B)     & 139 &  $-67$ & $-0.26$ \\
$8^+$   & (B)     &   25 &  $-60$ & $-$0.24& $6^+$ & (B)     & 143 &  $-66$ & $-0.26$ \\
$10^+$  & (B)     &   25 &  $-57$ & $-$0.22& $8^+$ & (B)     & 144 &  $-65$ & $-0.26$ \\
$12^+$  & (B)     &   23 &  $-52$ & $-$0.21& $10^+$& (B)     & 144 &  $-65$ & $-0.25$ \\
\hline
\multicolumn{3}{l}{$K=2$ band}   &      &      &       &         &     &     &      \\
$2^+$   & (A2';K=2)&   13 &   47 & 0.18 &       &         &     &     &      \\
$3^+$   & (A2';K=2)&      &      &      & $2^+$ & (A2';K=2)&  95 &  52 & 0.20 \\
$4^+$   & (A2';K=2)&   $-$9 &   61 & 0.24 & $2^+$ & (A2';K=2)&  38 &  57 & 0.22 \\
$5^+$   & (A2';K=2)&   $-$12 &   53 & 0.21 & $3^+$ & (A2';K=2)&  25 &  36 & 0.14 \\
$6^+$   & (A2';K=2)&  $-$18 &   62 & 0.24 & $4^+$ & (A2';K=2)&  83 &  60 & 0.23 \\
$7^+$   & (A2';K=2)&  $-$20 &   63 & 0.25 & $5^+$ & (A2';K=2)&  58 &  47 & 0.18 \\
$8^+$   & (A2';K=2)&  $-$22 &   63 & 0.25 & $6^+$ & (A2';K=2)& 104 &  61 & 0.24 \\
$9^+$   & (A2';K=2)&  $-$24 &   64 & 0.25 & $7^+$ & (A2';K=2)& 113 &  62 & 0.24 \\
$10^+$  & (A2';K=2)&  $-$22 &   57 & 0.22 & $8^+$ & (A2';K=2)&  96 &  56 & 0.22 \\
$11^+$  & (A2';K=2)&  $-$25 &   63 & 0.25 & $9^+$ & (A2';K=2)&  114 &  60 & 0.24 \\
$12^+$  & (A2';K=2)&  $-$28 &   67 & 0.26 & $10^+$& (A2';K=2)&  99 &  55 & 0.22 \\
\end{tabular}
\end{center}
\end{table}

\begin{table}
\caption{ \label{tab:be2np} Electric quadrupole moments $Q$ ($e$ 
fm$^2$) and $B(E2)$ 
($e^2$fm$^4$) for the negative-parity states 
calculated with the interaction(i). 
The extracted intrinsic quadrupole moments (fm$^2$) and deformation
parameters are also listed. The definitions of $Q^{t,s}_0$ and $\beta^{t,s}$
are explained in the text.
The labels for the bands are the same as those in 
Fig.\protect\ref{fig:level}.}
\begin{center}
\begin{tabular}{cccccccccc}
& & & & & \multicolumn {3}{l}{tansition to}& & \\
$J^\pi$ & band    & $Q$    & $Q^s_0$ & $\beta^s$ & final & band    & $B(E2)$ & $Q^t_0$ 
& $\beta^t$ \\
\hline
\multicolumn{3}{l}{negative-parity bands}   &      &      &       &         &     &     &      \\
\hline
$3^-$   & (C1)    &   -9 &   27 & 0.11 &       &         &     &     &      \\
$5^-$   & (C1)    &  -10 &   25 & 0.10 & $3^-$ & (C1)    &  25 &  29 & 0.11 \\
\hline
\multicolumn{3}{l}{prolate}   &      &      &       &         &     &     &      \\
$1^-$   & (C2)    &  -12 &   59 & 0.23 &       &         &     &     &      \\
$3^-$   & (C2)    &  -29 &   88 & 0.35 & $1^-$ & (C2)    &  89 &  59 & 0.23 \\
$3^-$   & (C2)    &      &      &      & $1^-$ & (C3)    & 133 &  72 & 0.28 \\
$5^-$   & (C2)    &  -31 &   81 & 0.32 & $3^-$ & (C2)    & 209 &  83 & 0.33 \\
$7^-$   & (C2)    &  -22 &   53 & 0.21 & $5^-$ & (C2)    &  76 &  49 & 0.19 \\
$7^-$   & (C2)    &      &      &      & $5^-$ & (C1)    &  26 &  28 & 0.11 \\
\hline
\multicolumn{3}{l}{large prolate}   &      &      &       &         &     &     &      \\
$1^-$   & (C3)    &  -20 &  100 & 0.39 &       &         &     &     &      \\
$3^-$   & (C3)    &  -28 &   85 & 0.34 & $1^-$ & (C3)    & 150 &  76 & 0.30 \\
$5^-$   & (C3)    &  -34 &   88 & 0.35 & $3^-$ & (C3)    & 216 &  85 & 0.33 \\
$7^-$   & (C3)    &  -40 &   98 & 0.38 & $5^-$ & (C3)    & 178 &  74 & 0.29 \\
$7^-$   & (C3)    &      &      &      & $5^-$ & (C2)    & 150 &  68 & 0.27 \\
\end{tabular}
\end{center}
\end{table}

\begin{table}
\caption{ \label{tab:be2-18} Electric quadrupole moments $Q$ ($e$ fm$^2$) and $B(E2)$ 
($e^2$fm$^4$) calculated with the interaction(ii). 
The extracted intrinsic quadrupole moments (fm$^2$) and deformation
parameters are also listed. 
The definitions of $Q^{t,s}_0$ and $\beta^{t,s}$
are explained in the text. The labels for the bands correspond to those in 
Fig.\protect\ref{fig:band-2}.}
\begin{center}
\begin{tabular}{cccccccccc}
& & & & & \multicolumn {3}{l}{tansition to}& & \\
$J^\pi$ & band    & $Q$    & $Q^s_0$ & $\beta^s$ & final & band    & $B(E2)$ & $Q^t_0$ 
& $\beta^t$ \\
\hline
$2^+$ & (A1)    &   $-$8 &   29 &  0.11 & $0^+$ & (A2)    &   27 &    37 & 0.14\\
$4^+$ & (A1)    &   $-$2 &    7 &  0.03 & $2^+$ & (A1)    &    9 &    18 & 0.07\\
$6^+$ & (A1)    &   $-$7 &   18 &  0.07 & $4^+$ & (A1)    &    2 &     8 & 0.03\\
\hline
\multicolumn{3}{l}{prolate} &      &      &       &         &     &     &      \\
$2^+$ & (A2)    &  $-$14 &   48 &  0.19 & $0^+$ & (A2)    &   48 &    49 & 0.19\\
$4^+$ & (A2)    &  $-$21 &   57 &  0.22 & $2^+$ & (A2)    &   60 &    46 & 0.18\\
$6^+$ & (A2)    &  $-$21 &   53 &  0.21 & $4^+$ & (A2)    &   89 &    53 & 0.21\\
$8^+$ & (A2)    &  $-$23 &   54 &  0.21 & $6^+$ & (A2)    &   88 &    52 & 0.20\\
$10^+$& (A2)    &  $-$29 &   67 &  0.26 & $8^+$ & (A2)    &  121 &    60 & 0.23\\
$12^+$& (A2)    &  $-$32 &   72 &  0.28 & $10^+$& (A2)    &  161 &    69 & 0.27\\
\hline
\multicolumn{3}{l}{superdeformation} &      &      &       &         &     &     &      \\
$2^+$ & (A3)    &  $-$40 &  138 &  0.54 & $0^+$ & (A3)    &  382 &   139 & 0.54\\
$4^+$ & (A3)    &  $-$49 &  135 &  0.53 & $2^+$ & (A3)    &  524 &   136 & 0.53\\
$6^+$ & (A3)    &  $-$52 &  129 &  0.51 & $4^+$ & (A3)    &  528 &   130 & 0.51\\
$8^+$ & (A3)    &  $-$49 &  116 &  0.45 & $6^+$ & (A3)    &  476 &   120 & 0.47\\
$10^+$& (A3)    &  $-$49 &  113 &  0.44 & $8^+$ & (A3)    &  363 &   104 & 0.41\\
$12^+$& (A3)    &  $-$54 &  120 &  0.47 & $10^+$& (A3)    &  383 &   106 & 0.41\\
\hline
\multicolumn{3}{l}{oblate} &      &      &       &         &     &     &      \\
$2^+$ & (B)     &   20 &  $-$71 & $-$0.28 & $0^+$ & (B)     &  102 &    $-$72 & $-$0.28\\
$4^+$ & (B)     &   25 &  $-$68 & $-$0.27 & $2^+$ & (B)     &  145 &    $-$71 & $-$0.28\\
$6^+$ & (B)     &   26 &  $-$66 & $-$0.26 & $4^+$ & (B)     &  157 &    $-$71 & $-$0.28\\
$8^+$ & (B)     &   26 &  $-$62 & $-$0.24 & $6^+$ & (B)     &  161 &    $-$70 & $-$0.27\\
$10^+$& (B)     &   23 &  $-$52 & $-$0.21 & $8^+$ & (B)     &  155 &    $-$68 & $-$0.27\\
\hline
\multicolumn{3}{l}{$\alpha$-cluster} &      &      &       &         &     &     &      \\
$2^+$ & (D)     &  $-$29 &  101 &  0.40 & $0^+$ & (D)     &  195 &    99 & 0.39\\
$4^+$ & (D)     &  $-$34 &   92 &  0.36 & $2^+$ & (D)     &  289 &   101 & 0.40\\
$6^+$ & (D)     &  $-$41 &  103 &  0.41 & $4^+$ & (D)     &  309 &    99 & 0.39\\
$8^+$ & (D)     &  $-$27 &   63 &  0.25 & $6^+$ & (D)     &  271 &    91 & 0.36\\
$10^+$& (D)     &  $-$26 &   61 &  0.24 & $8^+$ & (D)     &  420 &   112 & 0.44\\
$12^+$& (D)     &  $-$28 &   64 &  0.25 & $10^+$& (D)     &  440 &   113 & 0.44\\
\hline
\multicolumn{3}{l}{$K=2$:$\alpha$-cluster} &      &      &       &         &     &     &      \\
$2^+$ &  (D';K=2)&   31 &  110 &  0.43 & $0^+$ & (D';K=2)&      &       &     \\
$3^+$ &  (D';K=2)&      &      &       & $2^+$ & (D';K=2)&  390 &   105 & 0.41\\
$4^+$ &  (D';K=2)&  $-$20 &  139 &  0.55 & $2^+$ & (D';K=2)&  129 &   104 & 0.41\\
$5^+$ &  (D';K=2)&  $-$24 &  106 &  0.41 & $3^+$ & (D';K=2)&  216 &   107 & 0.42\\
$6^+$ &  (D';K=2)&  $-$30 &  104 &  0.41 & $4^+$ & (D';K=2)&  202 &    93 & 0.37\\
$7^+$ &  (D';K=2)&  $-$33 &  101 &  0.40 & $5^+$ & (D';K=2)&  237 &    95 & 0.37\\
$8^+$ &  (D';K=2)&  $-$50 &  143 &  0.56 & $6^+$ & (D';K=2)&  152 &    73 & 0.29\\
$9^+$ &  (D';K=2)&  $-$35 &  94 &  0.37 & $7^+$ & (D';K=2)&  267 &    95 & 0.37\\
$10^+$&  (D';K=2)&  $-$59 &  152 &  0.59 & $8^+$ & (D';K=2)&  250 &    90 & 0.35\\
\end{tabular}
\end{center}
\end{table}

\begin{table} 
\caption{\label{tab:Q2t} 
Intrinsic quadrupole moments (fm$^2$) of the normal-deformed band and
the superdeformed band extracted from 
$B(E2)$ with the definition in Eq.(\protect\ref{eq:Q2t}).
The present AMD results for the normal-deformed band and the superdeformed band
are those for the bands (A2) and (A3) with the interaction(i),
respectively. 
The values in parensis are those for the assignment of the
$8^+, 10^+, 12^+$ states in the band (A2) to the superdeformed states.
The shell model calculations by Poves {\em et al.} \protect\cite{Poves04},
and the GCM+HFBCS calculations by Bender {\em et al.}
\protect\cite{Bender03} are also listed. The experimental data are those 
taken from $B(E2)$ data in 
Ref. \protect\cite{nucldata},  and $Q^t_0$ values determined by
Doppler-shift analysis in Refs. \protect\cite{Ideguchi01} and 
\protect\cite{Chiara03}. $^a$For the values of
 Refs. \protect\cite{Ideguchi01} and 
\protect\cite{Chiara03}, a single $Q^t_0$ value is 
assumed for the entire transitions in each of the ``band 2'' or the 
``band 1''.}
\begin{center}
\begin{tabular}{cccccccccc}
$J_i$&	AMD &	GCM+HFBCS&	shell model&	
Nucl. data.\protect\cite{nucldata}&	Ideguchi {\em et al.}
\protect\cite{Ideguchi01}&	Chiara {\em et al.}\protect\cite{Chiara03}\\
\hline
ND & (A2) &&& &band 2& band 2\\
$2^+$&	57 &	75.2	&&	1.1$\pm 0.1\times 10^2$ &	
74$\pm 14^a$ \\
$4^+$&	64 &	23.9	&&	1.3$\pm0.2\times 10^2$	&74$\pm 14^a$	\\	
$6^+$&	64 &	77.4	&&			&74$\pm 14^a$	\\
$8^+$&	78 &		&&			&74$\pm 14^a$	\\
$10^+$&	105 &		&&			&74$\pm 14^a$	\\
$12^+$ &	112& 		&&			&74$\pm 14^a$	\\
\hline
SD& (A3) ((A2)) & & $8p$-$8h$ & & band 1& band 1\\
$2^+$	&129& 	133.9	&172&		&180 +39 $-$29$^a$ &	130$\pm 5^a$ \\
$4^+$	&129& 	97.6	&170&	1.4$\pm0.2\times 10^2$&180 +39 $-$29$^a$&	130$\pm 5^a$ \\ 		
$6^+$	&121& 	160.2	&167&		&	180 +39 $-$29$^a$ &	130$\pm 5^a$ \\
$8^+$	&97(70)&	157.9&	162&	&	180 +39 $-$29$^a$ &	130$\pm 5^a$ \\	
$10^+$	&81(105)&	161.2&	157&	&	180 +39 $-$29$^a$ &	130$\pm 5^a$ \\	
$12^+$	&78(112)	&	&160&	&	180 +39 $-$29$^a$ &	130$\pm 5^a$
\end{tabular}
\end{center}
\end{table}

\section{Discussion}\label{sec:discuss}

\subsection{Molecular bands}
Appearence of molecular resonances is
one of the important cluster aspects in $sd$-shell nuclei.
In the large prolate deformations of $^{40}$Ca, 
we find the trend of developed $^{12}$C+$^{28}$Si clustering 
which leads to the superdeformed band
as mentioned in the previous section. Therefore, it is
natural to expect possible existence of $^{12}$C+$^{28}$Si molecular bands  
and their link with the superdeformed band in $^{40}$Ca.

It has been known that molecular resonances appear in such systems as
$^{12}$C+$^{12}$C, $^{12}$C+$^{16}$O or $^{16}$O+$^{16}$O
\cite{Braun-Munzinger,Greiner95}.
In the microscopic calculations, there have been theoretical 
efforts to connect the molecular resonances 
with the low-lying deformed bands \cite{Kimura-S32,BAYE77,KATO85,Enyo-si28}.
We should stress that the recent AMD studies \cite{Kimura-S32,Enyo-si28} 
are the first theoretical 
works which can microscopically describe 
both the coexisting deformed states 
in low-energy region and the molecular resonances in the high-energy region
without relying on assumptions of constituent clusters.
In Ref.\cite{Kimura-S32}, 
the relation between the theoretically predicted 
superdeformed band in $^{32}$S and the $^{16}$O+$^{16}$O molecular 
bands has been studied. It has been suggested that
the superdeformed band with a considerable amount of the $^{16}$O+$^{16}$O 
component is regarded as the lowest nodal band
in a series of $^{16}$O+$^{16}$O molecular bands, 
while the excited $^{16}$O+$^{16}$O bands arise
due to the inter-cluster excitation and correspond to 
the observed molecular resonances.

\begin{figure}
\noindent
\epsfxsize=0.35\textwidth
\centerline{\epsffile{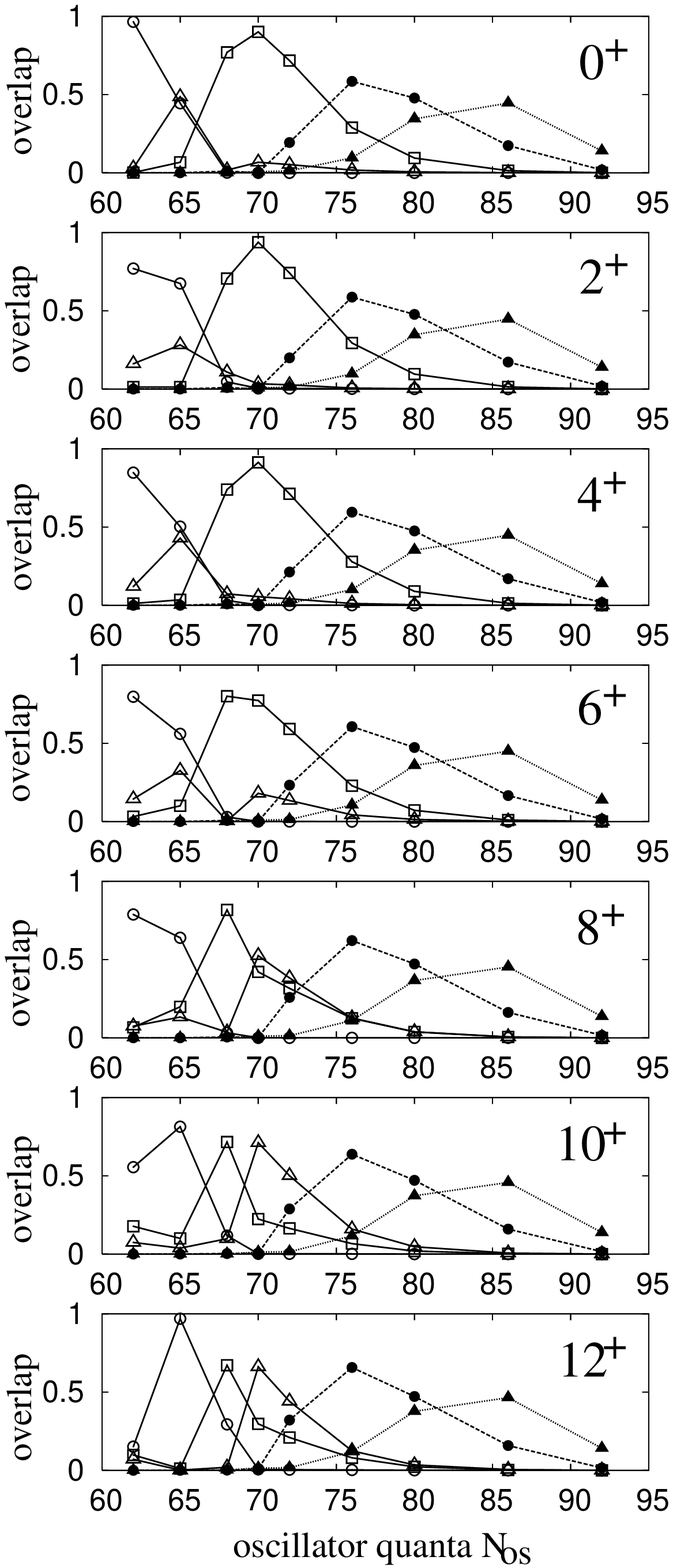}}
\caption{\label{fig:amp}
The overlap between the calculated excited states $\Psi_{n}^{J+}$
 and the prolate AMD states $P^{J}_{M0}\Phi^+_{\rm AMD}(N^{(+)}_{\rm os})$. 
The values $|\langle \Phi_{n}^{J+}|P^{J}_{M0}
\Phi^+_{\rm AMD}(N_{\rm os}^{(+)}) \rangle |^2$ 
for the states in the bands (A1), (A2), (A3), (A4) or (A5)
are shown by open circles, open triangles, open squares, filled circles
or filled triangles, respectively.
}
\end{figure}

In the present results of the case(i), we find
the excited rotational bands (A4) and (A5) as shown in Fig.\ref{fig:band}.
The former and the latter exist at about 15 MeV and 25 MeV 
higher excitation energies than that of the superdeformed band,
respectively. These bands appear due to the
superposition of the large prolate deformations with the $^{12}$C+$^{28}$Si
cluster structure. 
In Fig.\ref{fig:amp}, we show the overlap of the states($\Psi_n^{J+}$) 
in the prolate bands (A1), (A2), (A3), (A4) and (A5) with a single AMD state:
$|\langle\Psi_n^{J+}|P^J_{M0}\Phi^\pm_{\rm AMD}(N_{\rm os}^{(+)})
\rangle|^2$. 
The low-spin($J\le 4$) states in the superdeformed band is dominated by 
$P^J_{M0}\Phi^\pm_{\rm AMD}(N_{\rm os}=70^{(+)})$ with about 90\% overlap.
We should stress again that the $^{12}$C+$^{28}$Si-like clustering actually 
appears in the $\Phi_{\rm AMD}(N_{\rm os}=70^{(+)})$
(see Fig.\ref{fig:ca-dense}).
On the other hand,
the states in the excited band (A4) contain 
the AMD states with larger $N_{\rm os}$, and the components in the band (A5)
shift toward the further large $N_{\rm os}$ region.
Since the di-nuclear clustering 
develops further and two clusters go away from each other 
with the increase of $N_{\rm os}$, these excited bands (A4) and (A5)
should be described by the excitation of inter-cluster
motion. In fact, we calculate the oscillator quanta of the
subsystem consisting of the nucleons within each 
cluster($^{12}$C or $^{28}$Si), and find that  
each of the cluster is almost written by $0\hbar\omega$ configurations. 
In other words, the increase of $N_{\rm os}$
corresponds to just enlarging the inter-cluster distance,
and hence, the rotational bands (A4) and (A5) arise due to the
excitation of relative motion between $^{12}$C and $^{28}$Si clusters.
Moreover, it is suggested that the bands (A4) and (A5) may have tails in the
inter-cluster motion because the components of these bands are spread into 
the broader $N_{\rm os}$ ranges than that of the superdeformed band. 
These results suggest an interpretation that the superdeformed band (A3) 
and the higher bands (A4) and (A5) are regarded as 
a series of $^{12}$C+$^{28}$Si molecular bands.
Namely, the bands (A4) and (A5) appear 
as the higher nodal states built above the the superdeformed band.
These results have a good analogy to the feature of the superdeformed band 
and the $^{16}$O+$^{16}$O bands in 
$^{32}$S discussed in Ref.\cite{Kimura-S32}.

In the experimental side, the $^{12}$C+$^{28}$Si resonances 
have been observed in the elastic scattering data in the backward-angle region
\cite{Ost79,Braun-Munzinger}. The resonances at the energy 
$E_{\rm cm}=26.0$ and 30.2 MeV, which correspond to the excited states of 
$^{40}$Ca at $E=39$ and 44 MeV, are assigned to be $J=18$ states 
from the angular distributions. 
There are other experimental implications
of $^{12}$C+$^{28}$Si molecular states in the low-energy 
fusion cross section\cite{Racca83},
where distinct structures have been found in the energy region 
$E_{\rm cm}$=$20-30$ MeV above the $^{12}$C+$^{28}$Si threshold.
These are the candidates of the molecular resonances which might 
correspond to the $^{12}$C+$^{28}$Si molecular bands (A4) and (A5) obtained 
in the present results. In order to make 
further investigations of the molecular bands and give 
more quantitative discussions, it should be important to superpose 
a large number of the AMD states. It also may be effective to 
take into account the excitation inside the clusters as coupled channel
calculations for the description of the detailed band structure.

\subsection{single-particle orbits}

In an AMD wave function, the single-particle wave packets, $\varphi_i$,
in Eq.(\ref{eq:varphi}) are not orthogonal to each other. 
In order to study the mean-field character, it is useful to transform
them to the HF-like single-particle orbits $\{ \varphi^{\rm HF}_a \}$,
which are orthonormal to each other and form the total wave function
equivalent to the original AMD wave function.
We extracted the single-particle orbits $\{ \varphi^{\rm HF}_a \}$ from 
$\Phi_{\rm AMD}(N_{\rm os}^{(\pm)})$, as explained 
in Refs\cite{ENYOsup,Dote-be}, and analyze them. 

\begin{figure}
\noindent
\epsfxsize=0.9\textwidth
\centerline{\epsffile{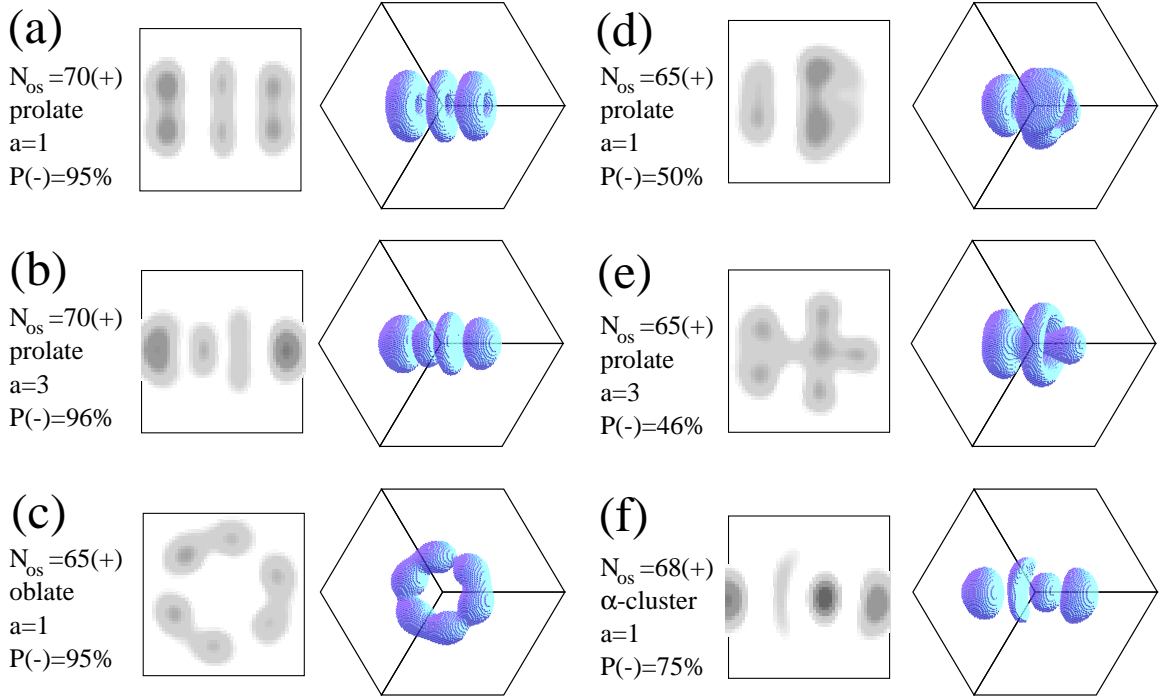}}
\caption{\label{fig:ca-single}
(Color online) <Density distributions of the HF-like single-particle orbits 
$\varphi^{\rm HF}_a$ for protons 
in $\Phi_{\rm AMD}(N_{\rm os}^{(\pm)})$.
The results are for the calculations with 
the interaction(i) except for the panel(f).
The first($\varphi^{\rm HF}_{a=1}$) and the third($\varphi^{\rm HF}_{a=3}$) 
highest proton orbits
in the the dominant 
component($\Phi_{\rm AMD}(N_{\rm os}^{(\pm)}=70^{(+)}$)
of the superdeformation
are shown in (a) and (b), respectively.
(c) The highest orbit, $\varphi^{\rm HF}_{a=1}$, in the oblate solution 
with $N_{\rm os}^{(\pm)}=65^{(+)}$. 
(d)The first($\varphi^{\rm HF}_{a=1}$) and 
(e)the third($\varphi^{\rm HF}_{a=3}$) orbits in the prolate solutions
with $N_{\rm os}^{(\pm)}=65^{(+)}$.
(f) The highest($\varphi^{\rm HF}_{a=1}$) orbit in the $\alpha$-cluster state
with $N_{\rm os}^{(\pm)}=68^{(+)}$ obtained in the results with 
the interaction(ii). 
In the left side, the intrinsic state is projected 
onto the $Y$-$Z$ plane, and the density is integrated along the $X$ axis,
where $X$, $Y$ and $Z$ axes are chosen as 
$\langle X^2\rangle\le \langle Y^2\rangle\le \langle Z^2\rangle$
and $\langle XY\rangle= \langle YZ\rangle= \langle ZX\rangle=0$.
The right figures are for the surface cut of the density.
The percentage $P(-)$ of the negative-parity component 
in each single-particle orbit is also listed. The box size is 10 fm.
}
\end{figure}

As mentioned before, the low-spin states of the superdeformed band
is dominated by the AMD state, 
$P^{J}_{M0}\Phi^+_{\rm AMD}(N_{\rm os}=70^{(+)})$.
We find that the intrinsic state, $\Phi_{\rm AMD}(N_{\rm os}=70^{(+)})$,
contains eight $fp$-like single-particle levels, which  
are the highest 4 levels among proton orbits and the highest 4 levels 
among the neutron orbits. Each of these levels has dominant 
parity-odd component as the ratio $P(-)\sim 95 \%$. 
The density distributions of these levels are
shown in Fig.\ref{fig:ca-single}(a) and (b). Hereafter, we use the label
$a$(=1,$\cdots$,20) of $\{ \varphi^{\rm HF}_a \}$ for
the $a$th highest single-particle levels in the proton orbits.
It is found that the properties of 
the eight valence levels in $\Phi_{\rm AMD}(N_{\rm os}=70^{(+)})$ 
are associated with the longitudinal $fp$ orbits which have 
$Y_{3,\pm1}$ and $Y_{3,0}$ angular dependent terms.
Therefore, we regard the superdeformed state as a ``$8p$-$8h$'' state, though
it is not equivalent to pure particle-hole states in the mean-field picture.
The origin of the parity asymmetric shape with $^{12}$C+$^{28}$Si cluster 
in $\Phi_{\rm AMD}(N_{\rm os}=70^{(+)})$ will be discussed later.

The prolate band (A2) contains the significant component of the
AMD state projected from the intrinsic state,
$\Phi_{\rm AMD}(N_{\rm os}=65^{(+)})$.
The valence orbits in this state are the eight levels which contain
50\% parity-odd components and 50\% parity-even ones as shown in 
figures \ref{fig:ca-single}(d) and (e). It means that the  
parity projected state $\Phi^+_{\rm AMD}(N_{\rm os}=65^{(+)})$ can be 
approximately described by the mixing of $2p$-$2h$, $4p$-$4h$ 
and $6p$-$6h$ configurations. We comment that 
$\Phi_{\rm AMD}(N_{\rm os}=65^{(+)})$ is not a local minimum
in the energy curve as a function of $N_{\rm os}$ nor $\beta$.

The oblate solution $\Phi_{\rm AMD}(N_{\rm os}=65^{(+)}_{\rm ob})$ 
does not mix with the prolate states, and independently 
forms the oblate band (B). This state can be convincingly described by 
a $4p$-$4h$ state because 4 valence nucleons occupy almost parity-odd
orbits with angular parts related to $Y_{3,\pm3}$. 
The property of a valence single-particle level of the four is shown in 
Fig. \ref{fig:ca-single}(c). 
The particle-hole nature is found also in 
the $\alpha$-cluster states $\Phi_{\rm AMD}(N_{\rm os}=66^{(+)}_\alpha)$
and $\Phi_{\rm AMD}(N_{\rm os}=68^{(+)}_\alpha)$ obtained by the 
interaction(ii). Namely, these states have a $4p$-$4h$ feature as
four valence neutrons
occupy $fp$-like orbits with 75\% negative-parity component in the
$\Phi_{\rm AMD}(N_{\rm os}=68^{(+)}_\alpha)$
as seen in Fig. \ref{fig:ca-single}(f).

\begin{figure}
\noindent
\epsfxsize=0.4\textwidth
\centerline{\epsffile{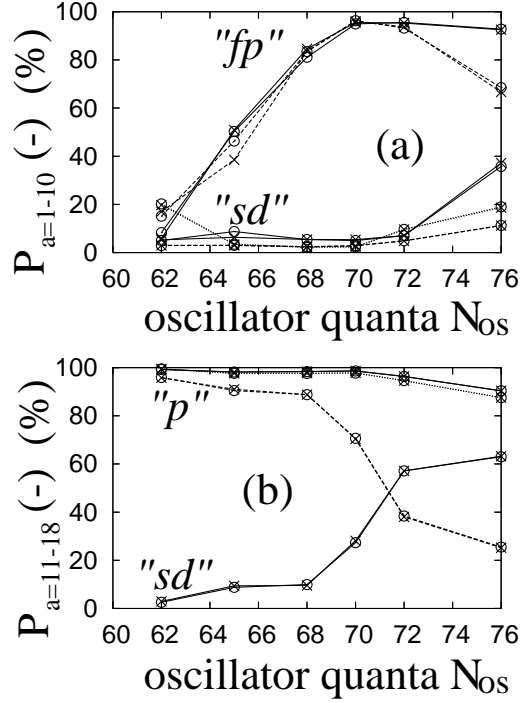}}
\caption{\label{fig:single}
The percentage $P(-)$ of the negative-parity component in each HF-like 
single-particle orbit for the protons in the prolate states
$\Phi_{\rm AMD}(N_{\rm os}^{(+)})$ obtained with the interaction(i).
The values $P(-)$ for the highest ten levels($a=1-10$) are shown in the
upper panel, and those for the orbits($a=11-18$) 
from the 11th to the 18th level  
are shown in the lower panel.}
\end{figure}

\begin{figure}
\noindent
\epsfxsize=0.3\textwidth
\centerline{\epsffile{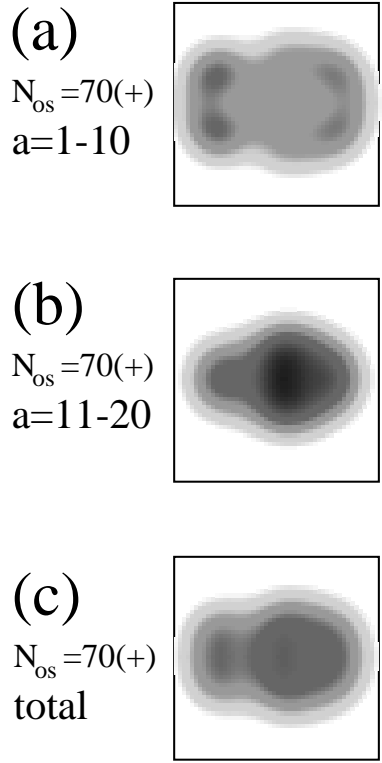}}
\caption{\label{fig:ca70-single}
Density distributions 
summed up (a) for the protons in the higher single-particle levels 
($\varphi^{\rm HF}_{a=1-10}$) and (b) for the protons 
in the lower single-particle levels($\varphi^{\rm HF}_{a=11-20}$)
in the intrinsic state of the 
superdeformation:$\Phi_{\rm AMD}(N_{\rm os}=70^{(+)})$. 
The total density is also shown in (c). The densities in (a) and (b) are
normalized by a factor 4.}
\end{figure}

Thus, the single-particle levels in the intrinsic states disclose
the many-particle many-hole characters in $^{40}$Ca. On the other hand,
the cluster aspect has been found in prolate states which have
parity-asymmetric shapes. It raises a question how the cluster aspect links
with the mean-field aspect. Here we examine the cluster aspect while paying attention to
parity asymmetry of the single-particle orbits. 
In Fig.\ref{fig:single}, we show the
ratio of negative-parity component of the single-particle proton orbits 
in the prolate states $\Phi_{\rm AMD}(N_{\rm os}^{(+)})$. 
The behavior of the neutron orbits
is almost the same as that of the proton orbits. 
At almost the spherical $N_{\rm os}=62$ region,
12 protons($\varphi^{\rm HF}_{a=1-12}$) occupy almost parity-even orbits and 
6 protons($\varphi^{\rm HF}_{a=13-18}$) are in almost parity-odd orbits,
which indicates the approximately $sd$-shell closed configuration 
of this state.
With the increase of $N_{\rm os}$, the parity-odd components 
in the highest four levels($\varphi^{\rm HF}_{a=1-4}$) increase
and these four levels become almost pure parity-odd orbits 
at $N_{\rm os}=70$. It means that four protons in the $sd$-shell 
are excited into the $fp$-shell.
In the state $\Phi_{\rm AMD}(N_{\rm os}^{(+)}=70^{(+)})$, 
each of the higher ten levels $\varphi^{\rm HF}_{a=1-10}$
has approximately good parity. 
Instead, the parity symmetry breaking in single-particle orbits
is rather remarkable in the 11th and 12th 
levels($\varphi^{\rm HF}_{a=11,12}$) which 
are originally the lowest
$sd$-orbits as shown in Fig.\ref{fig:single} (b).
In order to see the effect of the single-particle features
on the collective shape of the total system
$\Phi_{\rm AMD}(N_{\rm os}^{(+)}=70^{(+)})$, 
we show the density distribution 
summed up for the protons in the lower levels($\varphi^{\rm HF}_{a=11-20}$) 
and that for the protons in the higher levels($\varphi^{\rm HF}_{a=1-10}$) 
in Fig.\ref{fig:ca70-single}. 
The most striking thing is that 10 protons in the lower levels
form a parity asymmetric shape, which reminds us 
of the $\alpha$+$^{16}$O clustering in $^{20}$Ne
system. These results imply that the parity asymmetry in this state
dominantly originates not in the valence nucleons near the Fermi surface
but in the $\alpha$-cluster-like correlation 
in the core part. It is inconsistent with an mean-field picture 
that nucleons at the surface may contribute such an exotic
shape as the octupole deformation.
We here remind the reader that this 
state($\Phi_{\rm AMD}(N_{\rm os}^{(+)}=70^{(+)})$) 
is the dominant component of the superdeformation and has the 
$^{12}$C+$^{28}$Si-cluster structure(Fig.\ref{fig:ca70-single}(c)). 
The $^{12}$C+$^{28}$Si clustering is also seen in the aspect of 
molecular resonances. This seems to be inconsistent with 
possible $^{32}$S+2$\alpha$ clustering, which may be naively expected from 
the $8p$-$8h$ nature of the superdeformed state. 
By the analysis of single-particle orbits, we can interpret
the appearance of the $^{12}$C cluster as follows.
The origin of the $^{12}$C cluster in the superdeformation is 
understood by the correlation within 12 nucleons;
4 $sd$-shell nucleons with the $\alpha$ correlation and
the 8 valence nucleons in the $fp$-like orbits.
It leads to a new picture of the coexisting cluster and mean-field aspect.
Namely, the deeply bound nucleons 
compose the parity-asymmetric core $^{20}$Ne with the $\alpha$-cluster 
correlation, while the single-particle levels at the energy surface have
particle-hole nature.
In the further large $N_{\rm os}$ region, the particle-hole nature 
disappears because of the
spatial development of $^{12}$C+$^{28}$Si clustering.

\section{Summary}\label{sec:summary}

We studied deformed states in $^{40}$Ca with antisymmetrized
molecular dynamics(AMD) by using effective nuclear interactions 
which contain finite-range 2-body and 3-body forces.
In the framework of a constraint AMD method, we performed parity projection 
before energy variation, 
and total-angular-momentum projection after the energy variation. 
The obtained AMD wave functions are superposed to 
get better wave functions.

By analyzing the calculated results of the intrinsic states, 
level scheme, and $E2$ transition strengths,
it was found that the rotational bands with various deformations appear
in $^{40}$Ca as well as the spherical ground state.
Namely, above the spherical ground state, the rotational bands arise from 
the normal-deformed state and the superdeformed state
as well as the oblate state.
In the results with a weaker spin-orbit force, we also obtained 
the $\alpha$-cluster-like bands.
The present results of the superdeformed band reasonably reproduce 
the properties of the experimental superdeformed band build on the
$0^+_3$(5.21 MeV). On the other hand, we could not obtain 
a satisfactory description of the experimental deformed band 
build on the $0^+_2$(3.35 MeV). Possible assignment of this band 
is to the theoretical normal-deformed band,
the $\alpha$-cluster band, or admixture of them.  

In the anslysis of single-particle orbits, we found the particle-hole nature
of the superdeformed state, the oblate state and 
the $\alpha$-cluster state,
which are dominated by the $8p$-$8h$, $4p$-$4h$ and $4p$-$4h$ configurations,
respectively. On the other hand, the normal deformation in the
present results contains mixing of $2p$-$2h$, $4p$-$4h$ and $6p$-$6h$ 
configurations.

One of new findings in the present study is that the superdeformed state has
a parity asymmetric shape with $^{12}$C+$^{28}$Si-like cluster structure.
From the point of view of a single-particle picture, we discussed 
the relation between the cluster aspect and the many-particle 
many-hole aspect. We found that
the origin of the $^{12}$C clustering 
in the superdeformed state is understood by the correlation of 12 nucleons;
4 $sd$-shell nucleons with the $\alpha$-cluster correlation and
the 8 valence nucleons in the $fp$-like orbits.
It reveals the coexistence of cluster and mean-field aspect
in the superdeformed state. Namely, 
the $\alpha$-cluster correlation in deeply bound nucleons composing the core 
$^{20}$Ne plays an important role in the parity asymmetric shape, while
the particle-hole nature arises in the single-particle 
levels at the energy surface. 
The present result of the dominant $8p$-$8h$ feature in the superdeformation
seems to be consistent with the strong population 
of the members in the superdeformed 
band observed in 8-nucleon-transfer 
reactions;$^{32}$S($^{12}$C,$\alpha$)$^{40}$Ca\cite{Middleton72}.
However, it is inconsistent with the GCM+HFBCS calculations\cite{Bender03}, 
where neither the $4p$-$4h$ nor the $8p$-$8h$ HF solutions can be 
assigned to the bandheads of the normal-deformed band or superdeformed band.
One of the advantages of the
present framework is the parity projection before energy variation, which
is considered to be important in arising of the parity-asymmetric 
deformation. 
We think that one of the key properties which stabilize the
``$8p$-$8h$'' character in the superdeformed band is the 
$^{12}$C+$^{28}$Si clustering which causes a parity asymmetric shape.
Compared with the GCM+HFBCS calculations\cite{Bender03}, 
the higher correlations are ignored in the present calculations.
For example, the number of the basis in the superposition is ten at most, 
which is much smaller than the case of the GCM+HFBCS 
calculations\cite{Bender03}. Moreover, 
since we adopt the constraint on the total number of HO quanta 
in the energy variation, some basis are missed 
in terms of GCM with respect to the generator coordinate
$\beta_2$,  which are often used in the GCM+HF calculations.
The pairing correlations are also ignored. It is an remaining problem 
to check how the properties of the superdeformation 
are affected by introducing these effects.

Above the superdeformed band, we found possible higher rotational 
bands based on
$^{12}$C+$^{28}$Si-clustering in the large prolate deformations.
It is predicted that $^{12}$C+$^{28}$Si molecular bands may be 
built due to the excitation of the 
inter-cluster motion. The present results suggest that 
the superdeformed band and these higher molecular bands 
are regarded as a series of $^{12}$C+$^{28}$Si molecular bands.
This means that the $^{12}$C+$^{28}$Si cluster aspect is rather prominent 
in the superdeformed state than $^{32}$S+$2\alpha$ aspect, although 
the latter may be naively expected from the $8p$-$8h$ feature of this
state.
We also suggest the negative-parity 
bands caused by the parity asymmetric deformation.

In the present work, we use the stronger spin-orbit force than the 
nuclear interaction adopted in Ref.\cite{ENYO-3fb} to quantitatively 
reproduce the excitation energies of the superdeformed band 
and negative parity states. 
With the original weak spin-orbit force, the excitation energies 
of all the excited states in $^{40}$Ca are overestimated.
We would like to stress again that the intra-band properties such as 
energy spectra, $E2$ transitions and intrinsic structure 
of the superdeformed band are
not sensitive to choice of either interaction.
In order to solve the remaining problems, it may be 
necessary to introduce a suitable nuclear interaction 
and extended model wave functions.

\acknowledgments

The authors would like to thank Prof. H. Horiuchi and 
Dr. Taniguchi for many discussions.
They are also thankful to Dr. Ideguchi
and Dr. Inakura for valuable comments.
The computational calculations in this work were supported by the 
Supercomputer Projects of High Energy Accelerator Research Organization(KEK).
This work was supported by Japan Society for the Promotion of 
Science and a Grant-in-Aid for Scientific Research of the Japan
Ministry of Education, Science and Culture.
A part of the work was performed in the ``Research Project for Study of
Unstable Nuclei from Nuclear Cluster Aspects'' sponsored by
Institute of Physical and Chemical Research (RIKEN).

\end{document}